\documentclass[onecolumn]{aastex6} 

\usepackage{ulem}

\shorttitle{24 $\micron$ Survey of Young Stars}
\shortauthors{Meng et al. 2016}

\begin{document}
\title{The First 40 Million Years of Circumstellar Disk Evolution: the Signature of Terrestrial Planet Formation}

\author{Huan Y. A. Meng\altaffilmark{1}, George H. Rieke\altaffilmark{1,2}, Kate Y. L. Su\altaffilmark{1}, Andr\'{a}s G\'{a}sp\'{a}r\altaffilmark{1}}
\altaffiltext{1}{Steward Observatory, Department of Astronomy, University of Arizona, 933 North Cherry Avenue, Tucson, AZ 85721}
\altaffiltext{2}{Lunar and Planetary Laboratory, Department of Planetary Sciences, University of Arizona, 1629 East University Boulevard, Tucson, AZ 85721}
\email{hyameng@lpl.arizona.edu}

\begin{abstract}
We characterize the first 40 Myr of evolution of circumstellar disks through a unified study of the infrared properties of members of young clusters and associations with ages from 2 Myr up to $\sim$40 Myr: NGC 1333, NGC 1960, NGC 2232, NGC 2244, NGC 2362, NGC 2547, IC 348, IC 2395, IC 4665, Chamaeleon I, Orion OB1a and OB1b, Taurus, the $\beta$ Pictoris Moving Group, $\rho$ Ophiuchi, and the associations of Argus, Carina, Columba, Scorpius-Centaurus, and Tucana-Horologium. Our work features: 1.) a filtering technique to flag noisy backgrounds; 2.) a method based on the probability distribution of deflections, $P(D)$, to obtain statistically valid photometry for faint sources; and 3.) use of the evolutionary trend of transitional disks to constrain the overall behavior of bright disks. We find that the fraction of disks three or more times brighter than the stellar photospheres at 24 $\micron$ decays relatively slowly initially and then much more rapidly by $\sim$ 10 Myr. However, there is a continuing 
component until $\sim$ 35 Myr, probably due primarily to massive clouds of debris generated in giant impacts during the oligarchic/chaotic growth phases of terrestrial planets. If the contribution from primordial disks is excluded, the evolution of the incidence of these oligarchic/chaotic debris disks can be described empirically by a log-normal function with the peak at 12 - 20 Myr, including $\sim$ 13 \% of the original population, and with a post-peak mean duration of 10 - 20 Myr. 
\end{abstract}



\section{Introduction}

Most, if not all, young stars are born with circumstellar disks of gas and dust, which later give rise to planetary systems. Through a number of processes, protoplanetary disks lose most of their mass, including all primordial gas, within a few million years after their formation \citep{arm11,wil11,gorti16}. However, the formation of terrestrial planets through collisions and mergers of the remaining planetesimals can continue for tens of millions of years \citep[e.g.,][]{ken06,ken08,mor10,ste12,cha13,ray14} and can result in a renaissance of a dusty disk through planetesimal collisions \citep{genda15}. The pattern by which disks fade over their first $\sim$ 100 Myr provides clues to these processes, including dispersal by viscous accretion and photoevaporation, loss of dust grains by photon pressure and Poynting-Robertson drag, and through the consequences of the formation and migration of planets \citep[e.g.,][]{gorti16}. 

Over this entire time span, disks are marked by strong optical scattering and infrared and submillimeter emission by dust, which arises from primordial dust in the protoplanetary disks that are prominent for the first $\sim$ 5 - 10 Myr, and from second generation dust in the debris disks that come to the fore later. Different disk zones and particle sizes are probed in these dusty disks at different wavelengths, e.g., in optically thin (debris) disks sub-$\micron$ particles are most efficient at scattering, and small particles in the inner disk are also dominant in the near-infrared emission, while the far infrared reveals disk components at tens of AU  from their stars, and the submm is the realm of larger grains that mark the positions of the parent bodies whose collisions yield the dust seen at shorter wavelengths. Imaging in the optical with {\it HST} \citep[e.g.,][]{sch14} and recently in the near infrared with extreme AO systems \citep[e.g.,][]{mac14} has mapped disks in scattered light. Infrared observations with IRAS, ISO, WISE, {\it Spitzer}, {\it Herschel}, and in the brightest cases from the ground, have made substantial contributions to our understanding of disk evolution as seen in the emitting dust. Millimeter-wave facilities such as SCUBA, SMA, and ALMA have provided insights to the location of the parent bodies from which the grains seen at shorter wavelengths are generated.

A critical parameter to characterize disk dissipation and planet formation is the time evolution of these various disk components. Statistical studies, mostly in the wavebands from 2 $\micron$ to 8 $\micron$, have constrained the dissipation timescale of the inner zones of protoplanetary disks \citep[e.g.][]{hai01,har05,hil05,bri07,hern07a,wya08,mam09,pas10,wil11,rib14,clou14}. Starting with the pioneering study by \citet{hai01}, all of this work has converged on the conclusion that the fraction of disks above a given detection threshold decays exponentially with a decay timescale of $\sim$ 3 Myr. The 10 $\micron$ observations, though with poorer statistics, are consistent with this timescale \citep{aum91,mam04,hil05}. A similar timescale is deduced for the outermost disk regions from the general absence of sub-mm emission from systems that do not also have near infrared excesses \citep[e.g.,][]{and05,wil11}. The disk lifetime for higher mass stars is even shorter \citep{hai01,and05,mam09,fed10}.  

It has been assumed that disk dissipation in the giant planet forming zones that dominate disk emission between 20 and 40 $\mu$m follows a similar timeline \citep{pas10}. This is a demanding constraint on theories of giant planet formation \citep[e.g.][]{raf11}. However, the physical evolution in this region could well differ from that in the regions that dominate at shorter and longer wavelengths. Models of viscous accretion and photoevaporation predict that the mid-disk region, roughly coincident with the range for giant planet formation, should be the slowest to dissipate \citep{gorti15}. After the primordial disks have faded, the growth of oligarchics and their subsequent merger into terrestrial planets should also produce copious dust in dense debris disks.  The best available means to explore these processes is the sensitive measurements at 22 and 24 $\mu$m with WISE and {\it Spitzer}. However, such studies have been hampered by the need to extend them to relatively distant clusters and associations where the measurements do not reach the photospheres of most of the target stars.  The most ambitious effort \citep{rib14,rib15} assumes an exponential decay at 22 - 24 $\micron$ as has been found at the shorter wavelengths, but the uncertainties in the fitting parameters are large and the possibility of a different decay behavior is not excluded. For example, \citet{cur08a, cur08b} suggest that an exponential decay similar to that at shorter wavelengths  is followed by a second peak in disk incidence, attributed to the development of massive debris disks, whereas \citet{rib14,rib15} fit the behavior with a single exponential with a longer time constant than applies at the shorter wavelengths.

 In this paper, we explore disk evolution at 24 $\micron$, introducing three innovations to improve on previous studies: 1.) we introduce a digital filtering approach to eliminate from the study stars with potentially contaminated 24 $\mu$m data; 2.) we use a statistical method to constrain the number of faint sources below the detection limit for individual objects as a way to mitigate the inability of {\it Spitzer} to probe enough clusters to the deep limits desired; and 3.) we use the behavior of transitional disks, the final stage of evolution of protoplanetary disks prior to their dissipation, as an additional constraint on disk behavior.  

We have organized the paper as follows. In section 2, we describe our sample and age estimates. Appendices A and B provide more details on the clusters and associations from which we have drawn samples of stars for the study. Section 3 introduces the 24 $\mu$m data reduction and matching of infrared sources to cluster members. This section also introduces the digital filtering approach, which is described in more detail in Appendix C. In Section 4, we discuss the 24 $\mu$m excesses associated with disks and their time evolution, including use of statistical methods to extend the treatment to levels of disk emission below the individual-source detection limit. In this section, we also introduce the use of the transitional disk evolution as an additional constraint on the overall decay of disk incidence, requiring this decay to be non-exponential. Section 5 expands on these results, showing how the decay of 24 $\mu$m excesses is consistent with theoretical models of dissipation in the giant-planet-forming zones, and that there is a longer-lasting population of prominent disks possibly associated with giant collisions in the olligarchic/chaotic phase of terrestrial planet formation. 
Section 6 summarizes our conclusions.

\section{Cluster and Association Membership, Ages, and Completeness Limits}

\subsection{Membership}

For our study, we first identified young clusters and associations with a significant number of members and high-quality {\it Spitzer} observations. At the older ages, we supplemented this list with a small number of cases where WISE data have adequate sensitivity.  We proceeded to analyze only those clusters where high-quality membership lists were available from the literature, as discussed individually in Appendix~A. In the case of NGC 2244, we assembled existing material into a comprehensive membership list for this work (see Appendix~B). 

Our results would be undermined if the memberships had a significant number of stars with incorrectly identified infrared excesses. Therefore, we applied a digital filter to identify and eliminate cluster members on complex backgrounds that would significantly reduce the accuracy of the 24 $\mu$m photometry (and potentially yield false excesses at this wavelength). Our approach is similar to a selection method used previously on the basis of root median square deviation by e.g., \citet{meg12}, with some additional criteria.

\subsection{Ages}

 A recent revision of age estimates \citep[e.g.,][]{bell13,bell15} has resulted in larger values for most stars younger than 15 Myr than had been derived previously. We will distinguish this revised age scale from the traditional one. For ages $>$ 15 Myr, there is general agreement on ages as based on lithium absorption \citep{sod14}. We discuss the age estimates for each cluster/association 
in Appendix A. To 
obtain a homogeneous set of traditional ages, we gave priority to those listed in the review by Eric Mamajek \citep{mam09}. When necessary, we have assigned ages on the revised scale according to similarity
in published HR diagrams with clusters/associations with directly determined
ages. The two scales place the clusters/associations in a similar order but differ by up to a factor of two in the absolute values. We carry out our analysis on both scales to demonstrate that our conclusions are independent of which is used.

\subsection{Completeness}

 An inherent issue with studies of stellar clusters is the large range of distances and the resulting range of detection limits when converted, e.g., to stellar mass. The stellar clusters/assocations in this work span nearly two orders of magnitude in distance, from nearby moving groups with member stars within 20 pc to the farthest cluster at 1.5 kpc.

Two types of ``limits" are relevant in considering the effective range of our disk census. Given the membership list of a cluster or an association, first, we have a completeness in stellar mass, $M_{list,cpl}$, above which all member stars are likely already identified and included in the list. To determine the completeness limits of the photometry, we constructed luminosity functions in terms of the IRAC 3.6 $\micron$ magnitudes and estimated the limits as the magnitude where the functions turn over. Converted to masses, the limits are $\sim$0.5 M$_{\sun}$ at 300 pc and $/sim$ 1 M$_{\sun}$ at 1.3 kpc. Because of the shallow dependence of absolute K (and IRAC) magnitude on stellar mass, the resulting range in $M_{list,cpl}$ is modest and the trend nearly monotonically depends on cluster distance.


Second, depending on the distance and the depth of the MIPS image, there is a mass limit for each cluster to which the 24 $\mu$m measurements can detect stellar photospheres, $M_{det,ph}$. Although this mass is sometimes taken to limit 24 $\mu$m disk detections, studies focussing on large excesses are limited at a somewhat lower mass limit so long as $M_{list,cpl}$ $<$  $M_{det,ph}$. That is, in this case all of the disks can be detected to a lower mass limit than applies to the photospheres alone.  In section 4.1, we introduce a new method to estimate,  on a statistical basis, the disk brightnesses around member stars that are not individually detected at 24 $\micron$. With the new method, we can extend the mass limit for disk detection at our adopted excess level down to $\sim$ 1.2 $M_\odot$ even in the least favorable clusters, e.g., NGC 1960. 


\section{Data Processing}

For JHK and IRAC bands 1 - 4, we used publicly available/published measurements. However, we utilized approaches at 24 $\mu$m that required high quality new reductions. 

\subsection{Source Matching and Photometry}

We re-measured MIPS 24 $\mu$m photometry for all the clusters except for a few noted in Appendix A. We obtained the 24 $\mu$m data from the {\it Spitzer} Heritage Archive as summarized in Table 1. To extract the best possible photometric precision, throughout this work, we use only the mosaic images prepared by the MIPS Data Analysis Tool \citep{gor05} with post-pipeline processing to improve the flat fielding and to remove instrumental artifacts \citep{eng07}. The pixel scale in the final mosaics is 1.245\arcsec\ pixel$^{-1}$. This improved pipeline and calibration procedure produces high-quality mosaics with uniform photometric properties; the rms systematic photometric errors are well below 2\% \citep[See Appendix C.2 in][]{rie08}.

The open cluster IC 348 was observed in 8 epochs to search for variability \citep[e.g.,][]{flaherty12}. The depths of the images captured in each epoch are nearly identical. In this work we make measurements of stars on each image separately. The final photometry adopted is the average of all 8 measurements, and the error is their standard deviation. A fraction of Sco-Cen members were also observed in multiple epochs, where the same treatment is applied.

To associate 24 $\micron$ detections with those at shorter wavelengths, we start with the 2MASS (or [3.6]) positions for the member stars (whether the source is detected or not at 24 $\micron$), and convert them to image coordinates based on the standard world coordinate system (WCS) pointing information in the MIPS image headers. The astrometric error is expected to be on the sub-arcsecond level in 2MASS, and no worse than 1\arcsec\ in the MIPS images for stars with a high signal-to-noise (S/N) ratio. In identifying detections, a tolerance of 1.6 pixels (= 2.0\arcsec) is allowed to accommodate the lower S/N for many sources. This matching radius was selected to include virtually all of the {\it bona fide} matches but to minimize contamination by background sources; the study by \citet{bal16} demonstrates this performance.

Since the members are identified in the optical and near-infrared where stellar photospheres are brighter, many stars in the membership lists are not individually detected at 24 $\micron$. The sources that are detected individually make up our ``primary sample''. Classical PSF-fitting photometry was performed on the primary-sample cluster members with the DAOPHOT package in IRAF\footnote{IRAF is distributed by the National Optical Astronomy Observatories, which are operated by the Association of Universities for Research in Astronomy, Inc., under cooperative agreement with the National Science Foundation.}. Because the MIPS 24 $\mu$m PSF is very stable among data obtained at different times, instead of constructing a PSF from real stars in the images, we use a smoothed STinyTim simulated PSF \citep{kri06} for all the photometry  \citep{eng07}. 
We have found this procedure to give very accurate photometry. The PSF fit was optimized within the 2\arcsec\ positional uncertainty. To avoid interference from nearby bright sources or nebulosity, the PSF fitting was confined within fixed apertures with aperture correction factors determined from the simulated PSF. After some tests an aperture radius of four pixels (5$''$) was adopted for all clusters. To estimate the sky emission level under the PSF, sky annuli were determined cluster by cluster depending on the overall complexity of their respective sky backgrounds. As a general strategy we used small sky annuli to avoid serious background interference (see Table 1).

For each cluster, stars with low fluxes at 24 $\micron$ were inspected to determine the detection limits at which a stellar PSF became apparent in the images. Since the youngest clusters often have nebulous sky background, these limits vary significantly depending on the local sky appearance, typically corresponding to a nominal S/N $\sim$ 3-4 for areas with clear and stable background, and up to $\gtrsim$8 for clusters on very complicated background, like IC 348.  Table~\ref{list} lists the MIPS observations, as well as the photometric parameters used and the detection limits for each of the clusters. We consider stars with PSF-fitted flux densities lower than these image detection limits as undetected. However, some stars on complicated sky background with low PSF-fitting S/N ratios appear to have large measured flux densities in aperture photometry. The photometry in such areas is mostly unreliable, and is dealt with using a filtering process introduced in Appendix~\ref{filter}. This filtering is similar to that used by \citet{meg12} but with additional criteria to make it more selective. The basic idea of the filtering with both approaches is to eliminate the photometry on ``bad'' sky background. Such a position-based process is expected only to scale down the population of our sample, but to introduce no selection effects in terms of the statistical properties of the stars.
Cluster members on clean background but not detected at 24 $\micron$ comprise our ``secondary sample.'' For them, we obtain aperture photometry at the 2MASS positions with no recentering. Negative flux densities are allowed, since they could be natural results of statistical noise for non-detections and of unfavorable positions with respect to local bright regions on sky background.

Our photometry is summarized in Table 2.

\subsection{Determination of Relative Excesses at 24 $\micron$}

With the population of member stars (filtered to remove cases likely to be contaminated by complex backgrounds), we can determine the fraction of cluster and association members with 24 $\micron$ excesses. 
We define ``relative excess'', or the excess-to-photosphere flux ratio, as
\begin{equation}\label{eq1}
\epsilon = \frac{F_{24}^{disk}}{F_{24}^*} = \frac{F_{24}^{obs} - F_{24}^*}{F_{24}^*},
\end{equation}
where $F_{24}^{obs}$ is the observed flux density, $F_{24}^*$ is that expected from the stellar photosphere, 
and $F_{24}^{disk}$ is the deduced contribution from the disk. That is, $\epsilon$ is the 24 $\micron$ excess emission in units of photospheric brightness. Such relative measurements have the important advantage that they are not a strong function of stellar mass, and thus have been widely used in previous studies of circumstellar disks \citep[e.g.][]{wya08,gas09,sie10}.

The photospheric brightnesses at 24 $\micron$ are extrapolated from IRAC 3.6 $\micron$ photometry, except for the $\beta$ Pic Moving Group (BPMG) and the three subassociations of Sco-Cen where the extrapolations are based on the 2MASS $K_S$ photometry because it is more complete. We adopt 3.6 $\micron$ because data at longer wavelengths are more likely to have excess emission that is correlated to the 24 $\micron$ photometry and thus may not be indicative of the bare stellar photospheres, while the radiation at shorter wavelengths is likely to have suffered from more severe interstellar extinction that is nonuniform even within the same cluster. However, some young stars also have infrared excess at this wavelength \citep[e.g.][]{hai01}. These 3.6 $\micron$ relative excesses are generally much smaller than their 24 $\micron$ counterparts, and mostly fade out in a few Myr \citep[e.g.][]{hai01,wya08,mam09}. Therefore, we treat the 3.6 $\micron$ photometry differently for clusters younger than and older than 6 Myr old (on the revised scale, corresponding to $\sim$ 3 Myr on the traditional one). We discuss these two cases next.

\subsubsection{Clusters Younger than 6 Myr}

We include six clusters/associations younger than 6 Myr (3 Myr on the traditional scale), where we need to take account of possible excesses at 3.6 $\micron$: Taurus, $\rho$ Oph, Cha I, NGC 1333, NGC 2244, and IC 348. The first three are relatively large associations with reasonably well-behaved backgrounds; we adopt their 24 $\mu$m photometry directly from the literature, as introduced in Appendix~\ref{cluster_list}. The latter three, however, required new reductions and analyses of the MIPS data. We discuss our procedure for these three clusters in this section. 

 The first step in analyzing these results was to apply the digital filter (Appendix C) to eliminate all measurements likely to be contaminated significantly by diffuse emission. The second step is to estimate the extinction levels. The extinction, $A_V$, of IC 348 has been fitted for most member stars based on their spectral energy distribution (SED) in the optical and near infrared \citep{lad06}. For NGC 1333 where most stars (and all post-filtering members) are later than K4, we use the A$_V$ values in \citet{luh16}. For NGC 2244, we find no optical observations for most of the stars in our membership lists\footnote{\citet{par02} made visible $UBVI$ observations for the field of \objectname{NGC 2244}. However, out of 369 stars in our membership list, only 67 are identified with sources in their optical catalog with a matching radius of 3\arcsec.}, and we have to use $J$ vs $J - H$ to determine the extinction. For the extinction law, we adopt $A_V/A_{K_S} = 8.8$, $A_J/A_{K_S} = 2.5$, and $A_H/A_{K_S} = 1.55$ \citep{ind05,fla07}. In addition, we adopt $A_{\lambda} / A_{K_S} = 0.45$ for $\lambda$ between 3.6 $\micron$ and 24 $\micron$ \citep{ind05,fla07,cha09,fri11}. The approximation is valid because the extinction curve is almost flat from the IRAC bands to 24 $\micron$ \citep[e.g.,][]{fla07,cha09}. 

We then use the extinction-corrected measurements in a color-magnitude diagram to identify and estimate any excesses at 3.6 $\micron$, and correct our estimates of the excesses at 24 $\micron$ accordingly. 
For young stars with 3.6 $\micron$ excess, a conventional way to obtain the photospheric output is to examine a color-color diagram in the $J$, $H$, and 3.6 $\micron$ wavebands. However, color-color diagrams are not monotonic with stellar type in the JHK[3.6] range, so we turn to the color-magnitude diagram $J$ vs. $H-[3.6]$. Since all stars in a cluster have practically the same distance from us, their dereddened magnitudes are mutually comparable. The $J$ magnitude, as well as the $H - [3.6]$ color, are both monotonic with stellar mass. Hence, this color-magnitude diagram provides a valid relation to determine the 3.6 $\micron$ photospheric output.

There are a small number of stars ($\lesssim$ 10\% of the total in any cluster) for which we are unable to determine 3.6 $\mu$m excesses (e.g., stars without 2MASS detections). To reduce any resulting bias in our results, we estimate the photospheric levels for these stars assuming that they all have excesses that are the same as the average for the stars where we can determine the excesses individually. This may underestimate or overestimate the 3.6 $\micron$ photospheric output of individual stars. However, for the stars with individually estimated 3.6 $\micron$ photometry, the resulting correction to the excess at 24 $\micron$ is small, so the necessary scatter around the average for the remaining small number of stars will have little effect on our results. 

With these corrections, we derive the 24 $\micron$ flux density of the stellar photospheres, $F_{24}^*$, by the Rayleigh-Jeans relation from the 3.6 $\micron$ values. We use this approach because we do not have spectral types for the majority of the stars. We have found that the measured $[3.6] - [24]$ color is within 10\% of the Rayleigh-Jeans value for stars earlier than M-type, and the maximum offset for stars in our samples ($\sim$ M6) is $\sim$ 30 \% \citep{luh10}. In comparison, we will be analyzing excesses at the $\sim$ 300\% level (see \S 4.1), i.e., small errors in the extrapolated photospheric values have little effect on the derived excesses. The excess disk emission at 24 $\micron$, $F_{24}^{disk}$, is then obtained from equation~\ref{eq1}. 

\subsubsection{Clusters Older than 6 Myr}

The difference between considering the 3.6 $\micron$ excess and ignoring it becomes smaller as the clusters get older. For example, the corrections for 3.6 $\micron$ excesses for NGC 2362\footnote{5 Myr on the traditional scale, 12 Myr on the revised one} affect the disk frequency (defined in \S4) by only 0.7\%, lower than the Poisson counting uncertainty by a factor of $\sim$6.

For Sco-Cen (including Upper Sco) and the BPMG, rather than extrapolating from 3.6 $\mu$m  we utilized the 2MASS $K_S$ photometry because it is more complete. We first dereddened all the member stars with the extinction found in the literature, and then used the dereddened $K_S$ brightnesses of the stars to extrapolate to their 24 $\micron$ photospheric flux densities. This procedure should be accurate because the extinction levels are modest and the $K_S - [24]$ color depends only weakly on the spectral type for sources hotter than early M \citep{gau07}. Similar extrapolation of the photospheric output from $K$-band to 24 $\mu$m  has been found to be accurate at the $<$ 10\% level \citep{rie08}.

We have added the Pleiades based on the study of \citet{sie10}. With an age of 126 Myr \citep{sod14} at 136 pc \citep{mel14}, it is near the end of the expected age range of terrestrial planet formation around solar-like stars, and provides a reference for the end state of disk frequency in that era.

\section{Time Evolution of 24 $\micron$ Excess}

\subsection{``Disk'' Definition}

Disk frequency, defined as the fraction of stars with infrared excesses within a cluster, has been widely employed in previous studies as a statistical measurement of disk evolution. For this study, the disk frequency, $f$, at a relative excess threshold of $\epsilon_{lim}$ can be written as
\begin{equation}
f (\epsilon_{lim}) = \frac{N(\epsilon \geq \epsilon_{lim}) - N(\epsilon \leq -\epsilon_{lim})}{N},
\end{equation}
where $N(\epsilon)$ is the number of filtered member stars satisfying the limits indicated, and $N$ is the total filtered population of the cluster. The error in $f(\epsilon_{lim})$ is determined from the numbers of stars used in its calculation at any value of $\epsilon_{lim}$.


To define ``qualified disks,'' throughout this work we adopt a relative excess threshold of
\begin{equation}
\epsilon_{lim} = 3.
\end{equation}
Given the limited signal to noise, the distribution of detection levels may include some with excesses boosted above $\epsilon_{lim}$ by noise. Equation (2) allows us to correct for this effect on a statistical basis by subtracting those more negative than $- \epsilon_{lim}$. As discussed below, we use this approach to extend our survey to sources below the conservative single-source detection limit. The threshold $\epsilon_{lim} = 3$ roughly corresponds to fractional luminosity $L_{disk} / L_* \gtrsim 10^{-3}$, close to the maximum dust luminosity that a steady state debris disk is expected to reach ($L_{disk} / L_* \sim$ a few $10^{-4}$ in the age range considered in this work, e.g., ~\citet{wya07,ken08}). Debris disks typically have much lower fractional luminosities than primordial protoplanetary disks \citep[$L_{disk} / L_* \sim 1$, e.g.,][]{ken95}. Therefore, with this threshold most primordial disks should be counted, while only the most extreme debris disks could qualify \citep{bal09,mel10}.

Previous studies using disk counting have mostly been based only on stars bright enough to achieve good detections of their photospheres. Studies in the $JHKL$ and through the IRAC bands reach photospheric detection levels for solar-like stars out to $\sim$ 1 kpc, making the identification of members with excess infrared emission straightforward. However, such a method for identifying disks is less feasible at 24 $\micron$ except for the nearest associations. At 24 $\micron$ the MIPS detection limit is only sensitive enough for solar-like stars with $\epsilon_{lim} \ge 3$ out to $\sim$ 600 pc. 

 To extend this limit, we use an approach analogous to the probability of deflection, $P(D)$, methods originally developed for deep cosmological radio surveys to extract source counts below the threshold for reliable detection of individual objects \citep[e.g.,][]{ryle55,scheuer57}, but which now have found much broader use. For regions clear of background structure as determined by our filtering methods, we assume the intrinsic photometric errors on sky for each cluster are symmetric, i.e., away from individually detected sources the positive and negative error distributions are the same. This assumption is supported by \citet{pap04}, who showed that the negative side of the error distribution for MIPS 24 $\micron$ data is well behaved and does not include a significant number of spurious large values. Given the symmetric error distribution, we can use the random negative errors to estimate what fraction of the positive fluctuations in the cluster images exceed expectations for statistical errors. The excess of positive fluctuations should reveal the number of stars with real excesses on a statistical basis, i.e. not individually but through their influence on the spectrum of positive fluctuations. 

We apply this method to members of our ``secondary sample'', i.e., sources in clean sky but not detected individually. As described in Section 3.1, we obtained aperture photometry at the positions of these sources. We compare this photometry with the probability distribution of negative signals from similar photometry on the inverted images, i.e. with the negative error distribution. The excess of positive values at the positions of the sources is then a measure of their brightnesses on a statistical basis. 

Nonetheless, even this approach can only go into the noise by a modest amount before the errors become overwhelming. To illustrate the range of usefulness, we consider NGC 1960. Because stars fade as they evolve, NGC 1960, the oldest cluster in our sample farther than 1 kpc, should give us a worst-case scenario down to its $M_{list,cpl}$, the lowest-mass member star known, in this case about 1 M$_{\sun}$. The 1-$\sigma$ noise level of the 24 $\micron$ MIPS image of NGC 1960 is 0.072 mJy, compared to the faintest stellar photosphere at 0.020 mJy. A star with an excess above our threshold of $\epsilon_{lim} = 3$ would therefore have a 24 $\mu$m flux density of $\ge$ 0.060 mJy, that is only slightly below the 1-$\sigma$ detection limit. The limit to which our statistical method can be applied depends somewhat on the number of cluster members and of detections expected. In the limit where a large fraction of detections is expected, P(D) can reach well below the single-source detection limit before the uncertainties
resulting from its internal statistics rival those in the expected number of sources; however, if only a small fraction of detections is expected, P(D) is much less effective. With 132 identified members and detections at the level of 12.9\% for NGC 1960, valid estimates are returned down to about 1.5 $\sigma$ of the noise. Therefore, there is only a narrow range of flux density (0.06 to 0.11 mJy at 24 $\mu$m), and thus of stellar mass above $M_{list,cpl}$, over which we do not obtain valid estimates. This range is could only affect our estimate of the fraction of stars with excesses above $\epsilon_{lim} = 3$ if stars in this narrow range had an excess fraction dramatically different from the more massive stars, which is improbable.

\subsection{Decay functional form}

We show the overall decay of disk incidence with cluster/association age in Figures 1 and 2. With the exception of the low point for $\rho$ Oph, both figures indicate an almost constant incidence of qualified excesses for the initial phase of evolution (up to $\sim$ 6 Myr on the revised scale, $\sim$ 3 Myr on the traditional one). Although the decay of disk frequency is widely fitted with an exponential function, the behavior at 24 $\mu$m appears to be inconsistent with this behavior. Another constraint on the decay behavior is provided by transitional disks, which we show in this section argue strongly against exponential decay at 24 $\mu$m.

 \citet{bal16} show that among disks with strong 24 $\micron$ excesses, the incidence of transitional disks increases from $8.4 \pm 1.3$\% in stellar clusters between 1 and 6 Myr of age to $46 \pm 5$\% for those 9 - 12 Myr in age\footnote {both on the revised scale; the behavior on the traditional scale is identical but with the ages reduced by a factor of $\sim 2$}.
It is thought that the transitional stage is the final step before a primordial (Class II) disk loses its optically thick zones \citep{muz10,esp14} and thus its large 24 $\mu$m excess. \citet{cur09} have suggested homologous dissipation as another possible route. Referring to their classifications for NGC 2362, the members of this class fall equally into the weak transitional and modest-excess Class II categories in \citet{bal16}, and hence these disks also fall within that analysis. However, whether all of these systems will lose their excesses without ever assuming even a weak transitional form (as defined by \citet{bal16}) is not clear, since the mechanism(s) leading to the transitional stage are not well understood \citep{esp14}. In the following discussion we will make the common assumption that all disks {\it do} pass through the transitional stage \citep{muz10, esp14}. Any disks with a different evolutionary path would presumably also have a different disk fraction decay pattern.

\citet{bal16} show that the properties of the transitional disks do not appear to change significantly with cluster/association age. This behavior supports the view that all transitional disks are a similar phase that marks the loss of the inner disk material, independent of the age of the system. Therefore, the increase in relative numbers of transitional disks indicates that an increasing fraction of the remaining Class II sources are entering the transitional phase with increasing age. The mathematical definition of exponential decay would be that Class II stars lose that status at a constant rate relative to the remaining number. This is incompatible with the increase in the fraction of transitional disks relative to the combined number of transitional and Class II systems. Even if some systems skip the transitional stage, the large change in the proportion of transitional disks with age will strongly influence the overall disk fraction decay, in the direction just discussed. 

In addition, a simple model based on exponential decay only and the best-fit time constant cannot fit the transitional disk behavior; it predicts nearly as large (i.e., 2/3) a fraction of these systems in the young sample (1 - 6 Myr on the revised scale) as in the older one (9 - 12 Myr). Thus, the behavior of the transitional disks indicates that evolution of the 24 $\micron$ excesses deviates significantly from an exponential trend. Given the increase in the proportion of transitional disks with age, the deviation must be in the direction of the fading of Class II sources at larger ages being 
faster than the best-fit exponential, while the fading is slower than this exponential at smaller ages. 

\subsection{Fitting the Decay}

To discuss the non-exponential decay in the simplest possible way, we will use a heuristic model in which we assume that all stars start with Class II disks and that all disks pass from Class II to transitional to Class III, with the transitional phase having the same duration.  We can take duration of the transitional phase, $\tau_{tran}$, to be $\sim$ 1 - 1.5 Myr under the revised age scale\footnote{The duration of the transitional phase is estimated to be 0.5 -  1 Myr under the traditional age scale  \citep[e.g.,][]{muz10,clou14} .}. We take the 6 $\mu$m excesses to decay exponentially with a time scale of 5 Myr from \citet{clou14} and \citet{rib15}, the latter as recalculated for the revised age scale in this paper. We assume that the 24 $\mu$m excess decay is delayed by the transitional disk lifetime and thereafter decays exponentially with the same time constant as holds for the 6 $\mu$m excesses, i.e., 5 Myr. The result for $\tau_{tran}$ = 1 (1.5) Myr is a prediction of an average fraction of 0.69 (0.74) excesses in the 1 - 6 Myr range and a fraction of 0.17 (0.19) excesses in the 9 - 12 Myr range. These values are in satisfactory agreement with the data as summarized in Figure~\ref{1}a\footnote{If we adjust these results to the traditional age scale, the indicated transitional disk lifetime and hence the delay of the dissipation of the responsible disk component becomes less than 1 Myr.}.

However, the heuristic model does not include a sufficiently rapid increase in the relative number of transitional disks to match the observations \citep{bal16}. It therefore, at best, provides a very conservative upper limit to the number of primordial disks surviving past 10 Myr. A relatively simple modification to remedy this shortcoming is to assume that the relative rate at which systems enter the transitional stage from the pool of Class II sources increases linearly with time. Under the revised age scale, a coefficient of 0.04/Myr for the fraction of disks entering the transitional stage and a transitional disk lifetime of 1 Myr provide a reasonable fit to the data, with fractions of transitional disks of 9.7\% and 38\% respectively averaged over 0 - 6 Myr and over 9 - 12 Myr, compared with the observed values of 8.4 $\pm$ 1.3\% and 46 $\pm$ 5\%. This model also yields surviving fractions of disks averaged over 1 - 6 Myr of 0.86 and over 9 - 12 Myr of 0.18, similar to the values from the heuristic model. The evolution under this model is shown as the gray line in Figure 1a. The result is indicative, but not unique; a variety of assumptions about the rate of entering the transitional stage and its lifetime would be consistent with the constraints, but would all have generally similar decay behavior. The prediction for the loss of disks in the intermediate wavelength regime would also be modified from an exponential in this model, but given the scatter in the values for clusters as a function of age \citep{clou14,rib15}, this alternative to exponential decay is allowed.


\subsection{Evolution of Disk Frequency}

\begin{figure}
\center
\includegraphics[angle=+0,width=12cm]{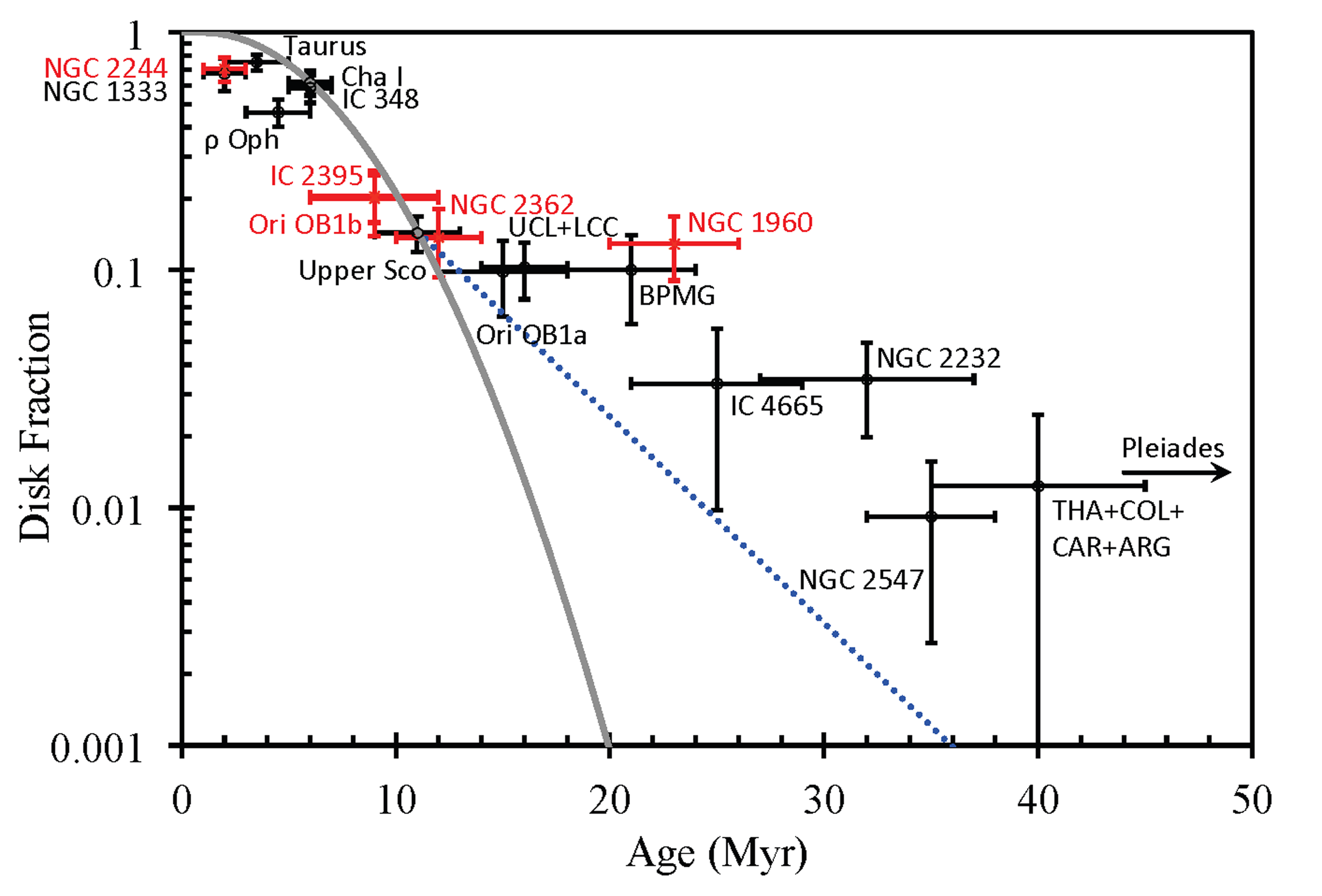}
\caption{ (a) Observed 24 $\micron$ disk frequencies as a function of stellar age on the revised scale, for excesses three times the photospheric emission ($\epsilon_{lim} = 3$). The black and red labels are stellar clusters and associations within and farther than 400 pc, respectively. The dotted blue line is the exponential decay of the heuristic model (exponential decay time constant of 5 Myr delayed by 1.5 Myr to allow for the transitional stage); it represents a conservative upper limit to the primordial disk population. The gray line shows a model taking account of the transitional disk behavior, with the best-fit linear dependence of entry rate to the transitional phase and a duration of 1 Myr for this phase (see Section 5.2). The disk frequency for the Pleiades is indicated by the arrow; the point would be at 126 Myr. The stellar clusters and associations between $\sim$15 and 35 Myr all have higher disk fractions than would be extrapolated from younger ages, representing a separate ``second component'' of disk evolution. 
\label{1}}
\end{figure}

We will preferably show the evolution of disk frequency on the revised age scale, since it provides a more demanding test of some important points we wish to make. When appropriate, we will show similar information on the traditional scale to illustrate that our conclusions hold equally well, or are stronger, using it. Figure~\ref{1}a shows the disk frequencies of the stellar clusters/associations in this study versus their ages (on the revised scale). The relevant values are listed in Table 3.  For reference, we show the prediction of our heuristic model, which does not account for the transitional disk behavior, and the fit taking account of the transitional disk behavior. For clarity, we do not show the heuristic model for ages $<$ 11 Myr. Both fits are consistent with the evidence that primordial disks older than 20 Myr are extremely rare \citep{wil11,ale14}. However, we find a significant fraction of stars with large excesses in this age range.

\setcounter{figure}{0}

\begin{figure}
\center
\includegraphics[angle=+0,width=12cm]{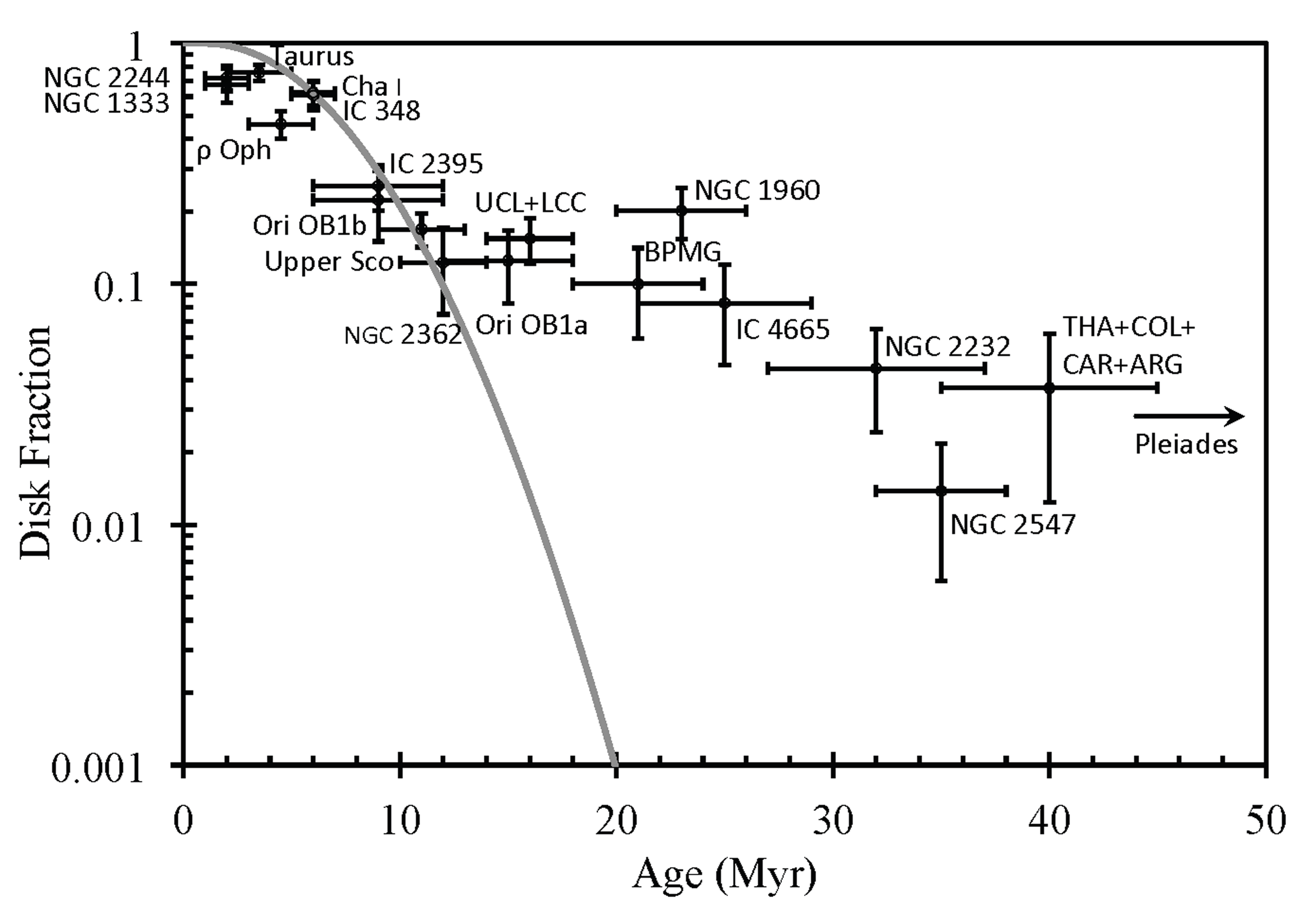}
\caption{ (b) Similar to Fgure 1a but with $\epsilon_{lim} = 2$. 
}
\end{figure}

\setcounter{figure}{0}

\begin{figure}
\center
\includegraphics[angle=+0,width=12cm]{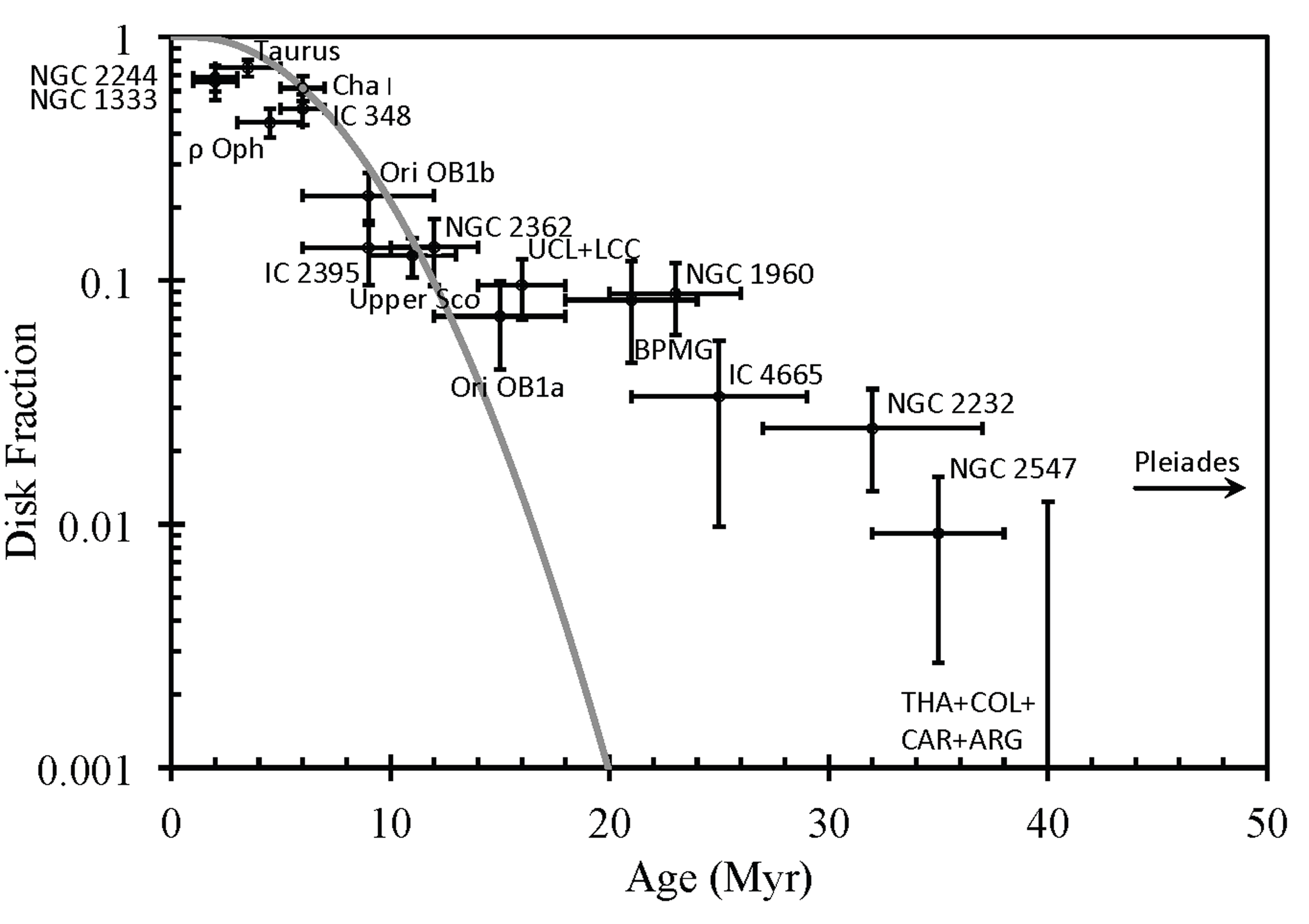}
\caption{(c) Similar to Figure 1a but with $\epsilon_{lim} = 4$.
}
\end{figure}

Figure 2 shows the same information as Figure 1 but on the traditional timescale. All the ages $>$ 15 Myr are expected to be the same in both cases \citep{sod14}, but the ages less than this value are compressed toward zero. As a result, the enhanced disk incidence around 20 Myr becomes very prominent. The behavior at ages $\le$ 3 Myr is similar to that $\le$ 6 Myr on the revised scale, although in the figure it is harder to see because of the compression.

\begin{figure}
\center
\includegraphics[angle=+0,width=12cm]{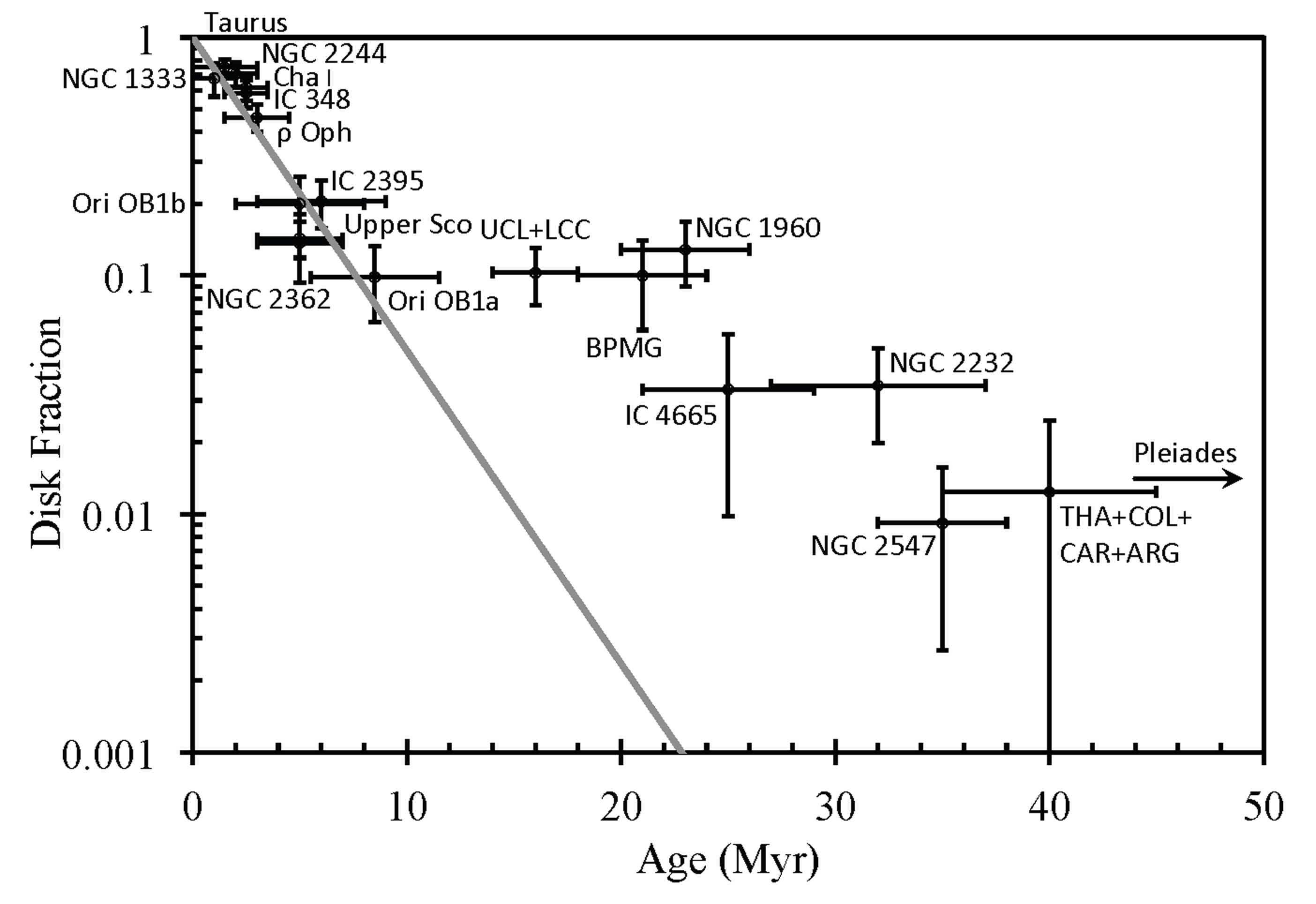}
\caption{Similar to Figure 1a, but with the clusters/associations on the traditional age scale and showing only an exponential fit to their decay. 
\label{2}}
\end{figure}

\subsection{Need for a second, slower component in disk decay}

Unfortunately, given the scatter among the results for the clusters/associations, their small number, and the uncertainties in age, it is not possible to derive a unique functional form for the decay of disk incidence from the 24 $\mu$m data. There is also a degeneracy between the duration of the transitional phase and the rate at which systems enter it that makes determination of a functional form from that information difficult. Nonetheless, Figures 1 and 2 show that there is a slowly decaying component that persists to $\sim$ 35 Myr above the level indicated by extrapolation of the behavior of the primordial disks in younger systems, after which the incidence drops to a level similar to that in the Pleiades. This component is apparent even relative to the heuristic model, which should overestimate the primordial disk incidence at these ages.

\subsection{Previous work}

For comparison with previous work, we begin by fitting the behavior at the shorter wavelengths with exponentials. Assuming an exponential decay of disk frequency from unity at $t = 0$, where $t$ is the cluster age \citep{mam09}, we fitted the evolution of the ten stellar clusters/associations of age 12 Myr or younger, since this age range gives us a relatively pure population of primordial disks. On the traditional age scale, we reproduce the values in \citet{rib15} well. Similar fits to the summary data in \citet{rib15} but with the revised ages give exponential decay time constants of 3.5 and 5.3 Myr respectively for the ``short ($<$ 6 $\mu$m)'' and ``intermediate (8 - 12 $\mu$m)'' wavelength behavior. The second value agrees well with the 5.75 Myr found by \citet{clou14} for IRAC excesses (out through 8 $\mu$m) using the revised age scale.

At 24 $\mu$m, the lifetime of qualified ($\epsilon_{lim} = $3) disks is found to be 6.6 $\pm$ 0.4 Myr, defined as the exponential time scale. We carried out a similar fit to the summary data in Table A.1 of \citet{rib15} but assigning updated ages to the clusters consistent with those in this work, obtaining 7.6 $\pm$ 1.3 Myr. The agreement is satisfactory, considering our different disk definitions and functional forms \footnote{Ribas et al. (2014; 2015) allow the disk fraction to decay from $<$ 1 to a baseline $>$ 0, whereas we start at 1 and decay to 0.}. On the traditional age scale, the exponential time scale is 3.2 $\pm$ 0.2 Myr.

Compared with the previous work \citep{rib14,clou14,rib15}, in addition to the different membership criteria, our approach is different in how ``disks'' are identified and counted. \citet{rib14,rib15} fit the SEDs of spectroscopically confirmed stars to obtain their expected photospheric fluxes at 24 $\micron$, and they count individual disks based on the significance of excess detection ($5\sigma$ above photospheric flux). Although they have tried to correct biases \citep{rib15}, this criterion is highly dependent on the sensitivity and depth of the observations (and hence the distance of the cluster), and also on the background behavior. Our work provides an independent test of their results, with the advantage that the disk frequencies should be less biased because we account for the net effect of individually undetected sources, and reject stars on noisy backgrounds regardless of the significance of their excess detections. It is therefore gratifying that the two methods find virtually identical exponential decay behavior at 24 $\micron$.  However, the exponential fits are largely empirical and do not take account of the actual stages thought to influence disk decay. Our heuristic model, in which the exponential decay is delayed by the time spent in the transitional stage, is a step toward a more realistic picture, and it yields a faster decay after the initial stages (beyond 10 Myr). Our model that takes account of the transitional disk behavior in more detail decays even faster.

\citet{rib14,rib15} also very sparsely sample the 11 - 40 Myr interval at 24 $\mu$m, which is represented only by 6 stars with excesses out of 43 in the BPMG. \citet{clou14} do not include stars older than 14 Myr. Thus, neither of these studies provides a good test of the disk behavior at 20 Myr and beyond. As shown in Figures~\ref{1} and~\ref{2}, the observed disk frequencies of the clusters older than $\sim$20 Myr are indeed higher than the extrapolated trends from the younger clusters, even assuming the very conservative heuristic model. For an extremely conservative estimate of the significance of the enhanced incidence of excesses, we apply a $\chi^2$ test to the clusters between 15 and 40 Myr (both inclusive) with the purely exponential decay ($\tau$ = 7 Myr) fitted for the first 12 Myr on the revised scale. We find that the probability that the disk frequency of the older clusters follows the same trend as the younger clusters is $<$ 10$^{-6}$, showing the necessity for a second, more slowly decaying disk component to fit the data. The probability would be substantially lower with the heuristic model, and lower still with the model incorporating the transitional disk behavior.


\subsection{Robustness of Results}

The evidence for the ``second component'' should not depend critically on our choice of the threshold, $\epsilon_{lim} = 3$. 
We tested the robustness of our results by redoing Figure~\ref{1}a under new thresholds $\epsilon_{lim} = 2$ and $\epsilon_{lim} = 4$. The outcomes are shown in Figures~\ref{1}b and c, respectively. The exponential decay timescales of disk frequency, with the clusters younger than 12 Myr, are $6.9 \pm 0.4$ Myr with $\epsilon_{lim} = 2$ and $6.3 \pm 0.4$ Myr with $\epsilon_{lim} = 4$, i.e. not significantly different. Similarly, regardless of the difference in disk definition, the disk frequency starts to level out in the 10-15 Myr age range and the ``second component'' remains prominent: the stellar groupings older than $\sim$15 Myr are all above the extrapolated trend. 

A potential risk in comparing the clusters over a wide range of distances arises from the different completeness limits in the membership lists. If the disk presence is correlated with spectral type, we may see different disk frequencies as a result of looking into different mass regimes. To address this issue, we take NGC 1960 as the reference for stellar mass, since it is the farthest cluster older than 20 Myr in our sample and appears to lie in the middle of the age of the possible ``second component'' in disk evolution. The faintest detected member in NGC 1960 has an extinction-corrected magnitude of $[3.6] = 14.2$, corresponding to $\sim$1.2 M$_{\sun}$. Therefore, we cut off the lower part of the membership lists of all nearer clusters and associations, making subsets of their members with this same mass limit\footnote{With the 1.2 M$_{\sun}$ cutoff and the general mass function \citep{kro01}, we estimate the average stellar masses of the stars in the subsets of all clusters are around 1.5 M$_{\sun}$.}. The relation between stellar mass and photospheric brightness is more isochrone-dependent toward younger ages \citep{hil04,hil08}. For the most robust results, we only retained the stellar clusters/associations older than 9 Myr. The mass cut-off is not applied to NGC 2362, whose distance and completeness limit are similar to those of NGC 1960, or to UCL and LCC, whose membership lists are already selected for F- and G-types. The result based on the trimmed membership lists is shown in Figure~\ref{2}, overplotted on the heuristic and transitional-disk-constrained decay lines obtained from the full membership lists, i.e., the same decay lines as in Figure~\ref{1}a. Although the member counts of some clusters have shrunk and thus enlarged the uncertainties, the disk frequencies of all clusters remain consistent within the errors with the results derived from the full membership lists.

To examine the behavior at lower masses, we show in Figure~\ref{4} the behavior for a mass range of 0.5 - 1.2 M$_\odot$, determined from clusters/associations within 400 pc. The results are very similar. The number of stars $\ge$ 20 Myr old with excesses above $\epsilon_{lim} = 3$ significantly exceeds the extrapolations from the behavior of primordial disks. To be specific, compared with the heuristic model and on the revised age scale, and assuming a Poisson distribution for the excess counts between 20 and 35 Myr, they have less than a 1\% probability of arising by chance. The likelihood of their arising by chance with the model including the constraint from the transitional disks is much lower. These arguments are only strengthened under the traditional age scale.
 
\begin{figure}
\center
\includegraphics[angle=+0,width=12cm]{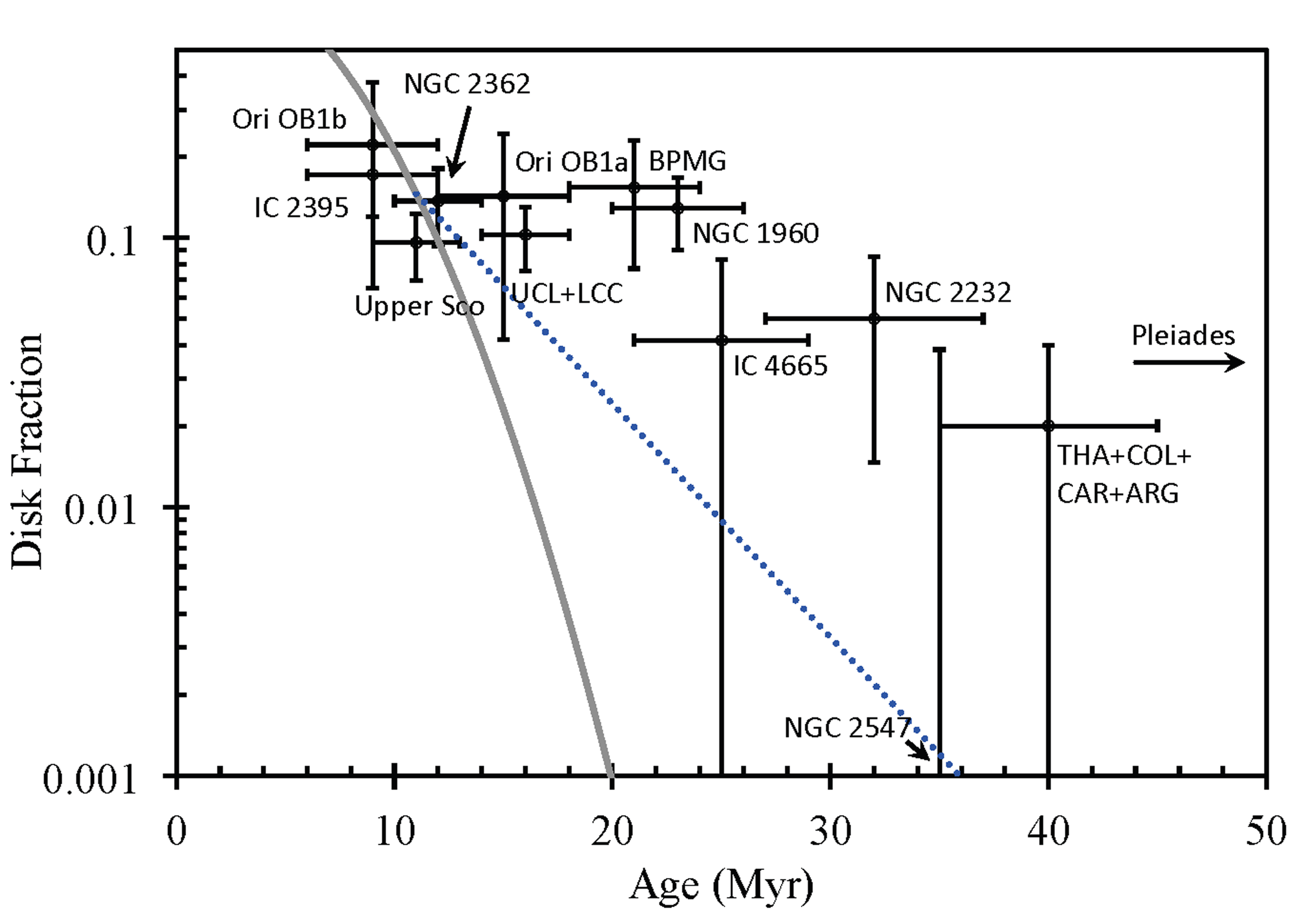}
\caption{Same as Figure~\ref{1}a, but only for stars $>$1.2 M$_{\sun}$. The gray line is the decay fitted with the full membership lists in Figure~\ref{1}a. The dotted blue line is the heuristic model that represents a conservative upper limit to the primordial disk population.
\label{3}}
\end{figure}

\begin{figure}
\center
\includegraphics[angle=+0,width=12cm]{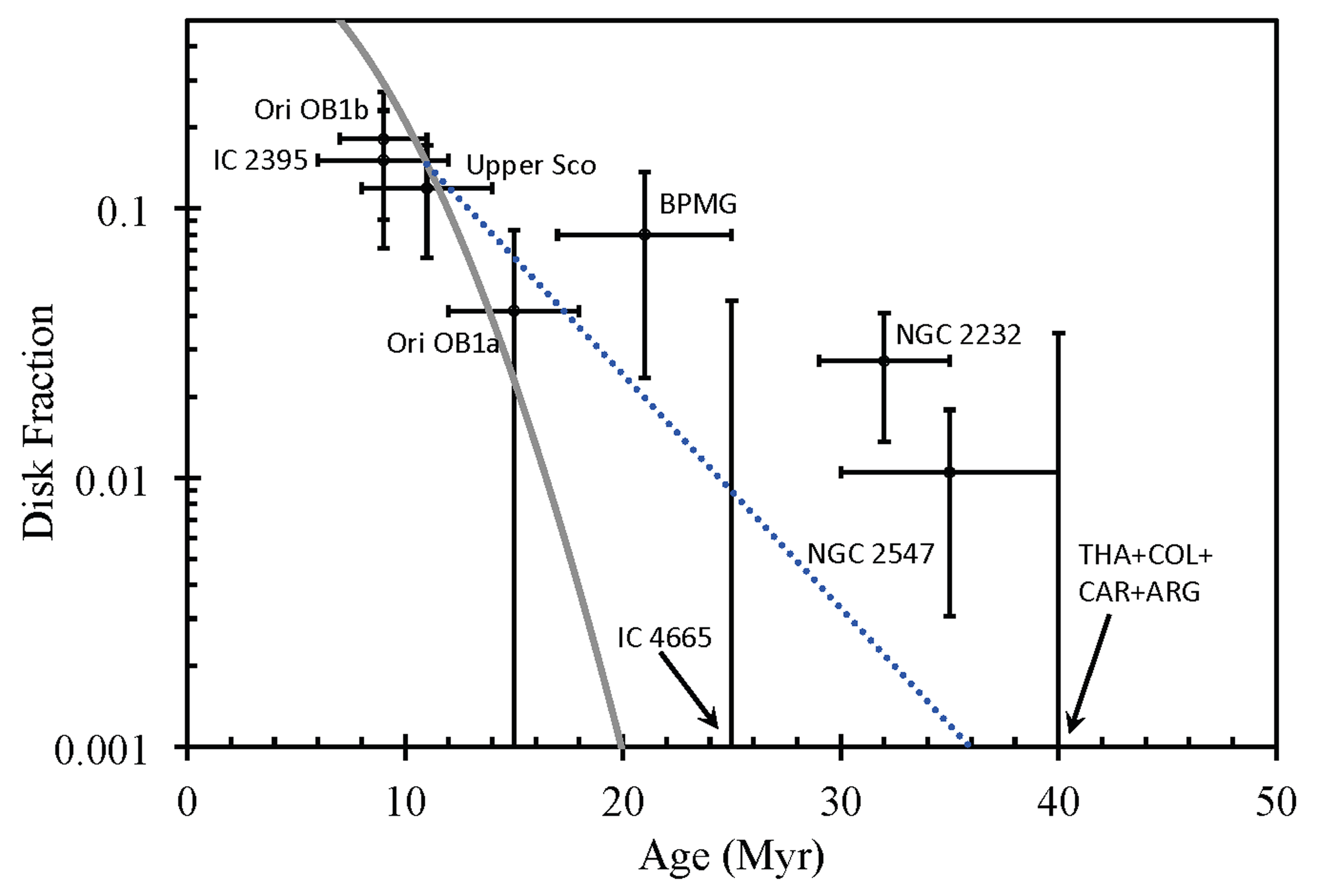}
\caption{Same as Figure~\ref{1}a, but only for stars between $0.5$ and $1.2 M_{\sun}$. The gray line is the decay fitted with the full membership lists in Figure~\ref{1}a. The dotted blue line is the heuristic model that represents a conservative upper limit to the primordial disk population. 
\label{4}}
\end{figure}

\section{Discussion}

\subsection{Lifetime of the 24 $\mu$m Protoplanetary Disk Component}

All of the fits (simple exponential, heuristic, and transitional-disk) show that the decay at 24 $\mu$m is significantly slower than at the intermediate infrared wavelengths ($8 - 12 \mu$m) defined by \citet{rib14} (e.g., from our fits to the summary in \citep{rib15} and on the revised age scale, from 5.3 to 7.6 Myr respectively at $8 - 12 \mu$m and 24 $\mu$m). This general behavior is predicted by photoevaporation models that show erosion from inside and outside with the most durable radial zones lying near those dominating the 24 $\mu$m emission \citep{gorti15,gorti16}. The theoretical prediction is for a delay at 24 $\mu$m only of order 0.5 Myr after the dissipation of the region emitting near 6 $\mu$m \citep{gorti15}. The difference in exponential decay time constants is significantly larger, but the agreement is much better with the two more realistic fits to the decay behavior. With them, the predicted $\sim$ 1 Myr delay (on the revised age scale) in the erosion of the disk component dominant at 24 $\mu$m compared with those seen at shorter wavelengths is consistent within a factor of two with the theoretical prediction \citep{gorti15}\footnote{If we use the traditional age scale, the behavior is even more consistent with the prediction.}. We conclude that our analysis supports theoretical work by \citet{gorti15,gorti16} that predicts that disks dissipate by photoevaporation from both the inside (near the star) and the outside. That is, the dissipation of the disk zone where giant planets form should proceed under this mechanism more slowly by about 0.5 - 1 Myr than that of the zones inside and outside this one.

\subsection{The ``Second Component'' of Disks Between 10 and 35 Myr}

We now discuss the nature of the systems responsible for the ``second component'' of prominent disks. We will quantify their behavior for stars more massive than 1.2 M$_\odot$, since this threshold gives us relatively complete samples even for the more distant clusters/associations. Given the difficulty in deriving a functional form for the decay of the protoplanetary disks, we have removed the primordial and transitional disks individually from the counts for the clusters/associations near 10 Myr (revised time scale), generally as identified in the papers providing the cluster membership lists\footnote{\citet{bal16, dah07, hern07b, car09, che11, reb08, smi12}}. The resulting counts for disks around $\geq$1.2 M$_{\sun}$ stars and from 9 to 40 Myr are shown in Figure 4 (the relevant fractions of excesses can be found in Table 3). We find that the evolution of the disk frequency in this age range can be described by a broad peak. If fitted empirically with a log-normal function and on the revised age scale, the peak frequency of these bright disks above the threshold of $\epsilon_{lim}=3$ is $12.3 \pm 2.9\%$ and occurs at $18.1 \pm 1.1$ Myr. Based on the assumption of a log-normal evolution, integrating from the age of the maximum incidence afterwards, the mean duration of the high level of activity of the qualified debris disks is $11.8 ^{+2.6} _{-3.3}$ Myr. Similarly, if we use the traditional age scale the fitted parameters are peak of $13.1 \pm 3.4\%$ at $11.8 \pm 1.2$ Myr, with a nominal post-peak lifetime of $16.0^{+6.1} _{-4.8}$ Myr. The errors are estimated through a Monte Carlo simulation of 10,000 runs distributed normally as indicated by the nominal error bars.

\begin{figure}
\center
\includegraphics[angle=+0,width=12cm]{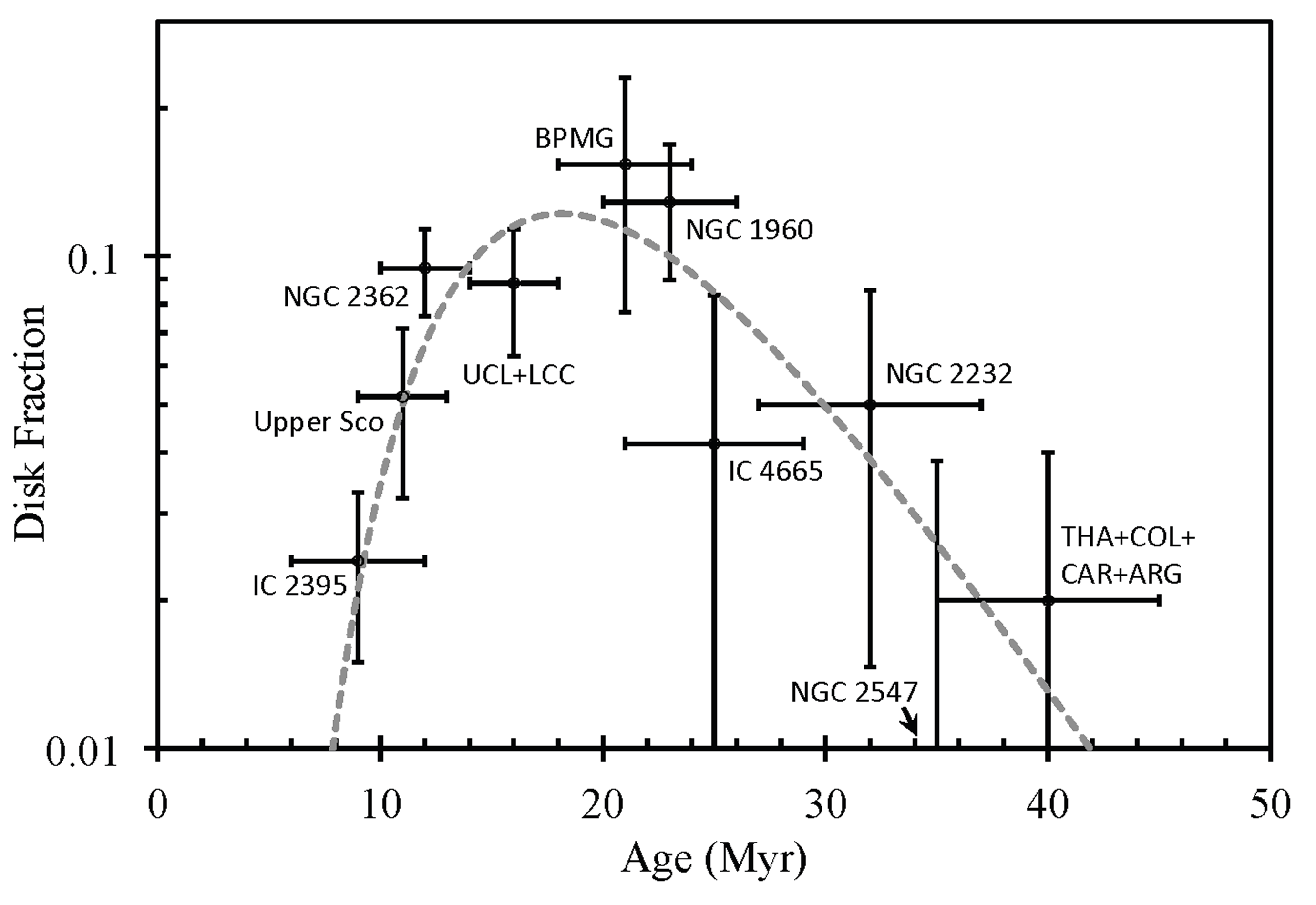}
\caption{Incidence of debris disks around stars $>$1.2 M$_{\sun}$, with the contribution from protoplanetary disks removed. The dashed line shows the best-fit log-normal function, with the peak of 12.3\% $\pm$ 2.9\% at 18.1 $\pm$ 1.1 Myr and a mean lifetime of $11.8 ^{+2.6} _{-3.3}$ Myr.
\label{5}}
\end{figure}

\begin{figure}
\center
\includegraphics[angle=+0,width=12cm]{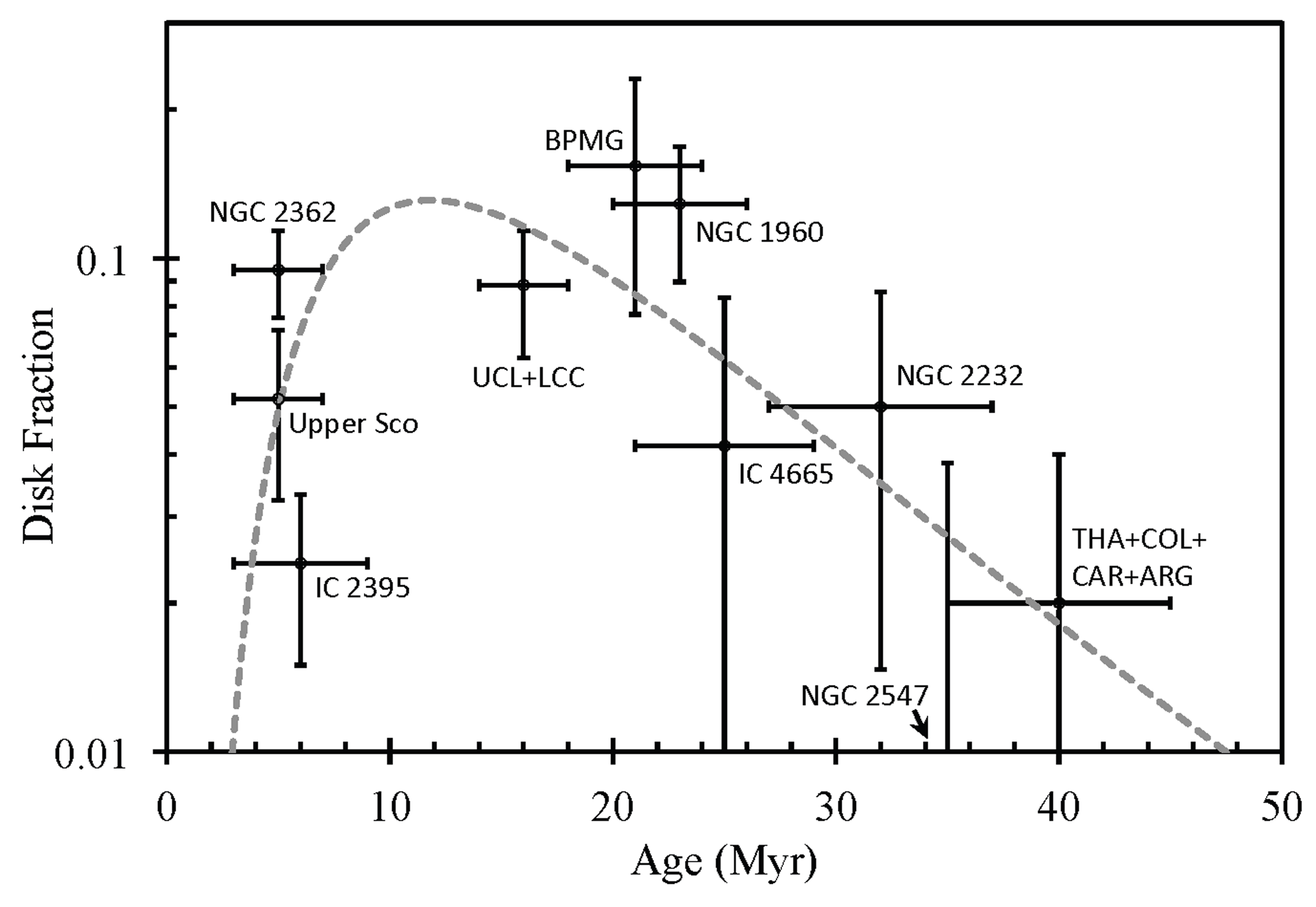}
\caption{Similar to Figure 4 but on the traditional age scale. The best-fitting log-normal function, has a peak of $13.1 \pm 3.4\%$ at $11.8 \pm 1.2$ Myr, with a nominal post-peak lifetime of $16.0^{+6.1} _{-4.8}$ Myr.
\label{6}}
\end{figure}

In the following sections we consider possible explanations for this component of disk behavior.

\subsubsection{Protoplanetary Disks Cannot Account for the Excesses at 15 - 35 Myr}

We first consider the possible persistence of protoplanetary disks into the 10-20 Myr range. \citet{car09} found 24 primordial disks and 2 Be stars in Upper Sco, while \citet{che11} found none. One of those 26 disks (HD 142184) is below our excess threshold. That is, there are 23 identified primordial disks in Upper Sco and above our excess threshold, including 6 with masses higher than 1.2 M$_{\sun}$. For NGC 2362, \citet{dah07} list 13 disks in their Table 3 with primordial-like SEDs and spectral signatures of accretion. The current membership lists of Ori OB1a and OB1b both have too few stars that are more massive than 1.2 M$_{\sun}$ to contribute to this discussion. Two detected disks in UCL and one in LCC are identified as primordial \citep{che11}. However, the one in LCC (HD 101088) is a rare case of a very low excess accretion disk that is fainter at 24 $\micron$ than our threshold \citep{bit10}. Since it is never counted as a ``disk'' under our definition, we only identify the two UCL primordial disks from the combination of UCL and LCC. In summary, there appears to be a minority population of protoplanetary disks up to $\sim$16 Myr.

We also investigated the nature of the individually detected disks in the older, 20-40 Myr age range. In the BPMG, all members are individually detected and there are 6 qualified ($\epsilon_{lim}=3$) disks: $\beta$ Pic (A6), V4046 Sgr (K5 + K7), HD 172555 (A7), $\eta$ Tel (A0), HD 181327 (F6), and HD 191089 (F5). The V4046 Sgr disk is accreting \citep[e.g.,][]{stemp04, gunther06} and classified as transitional \citep[also circumbinary, see][]{ros13}. However, the other five are bright debris disks. The other clusters in this age 
range are not known to contain accreting stars at a level that would qualify for our sample. This behavior is consistent with 
the general evidence that primordial disks do not survive up to this age range \citep{ale14} and therefore cannot account for the second disk component. 

\subsubsection{Steady-state Debris Disk Contribution is Small}

We next consider whether this peak in disk incidence is the initial starting point of the observed collisional steady-state decay of most known debris disks, such as that shown in \citet{gas13}.
During the initial build-up of dust in the systems while they are approaching the quasi-steady-state equilibrium, the infrared emission of a single system peaks. The peaks of various initial mass and spectral-type systems coincide roughly with the period of oligarchic planet formation. Therefore, simple steady-state decay could conceivably produce the observed peak in disk excess fractions.

\begin{figure}
\begin{center}
\center
\includegraphics[angle=+0,width=12cm]{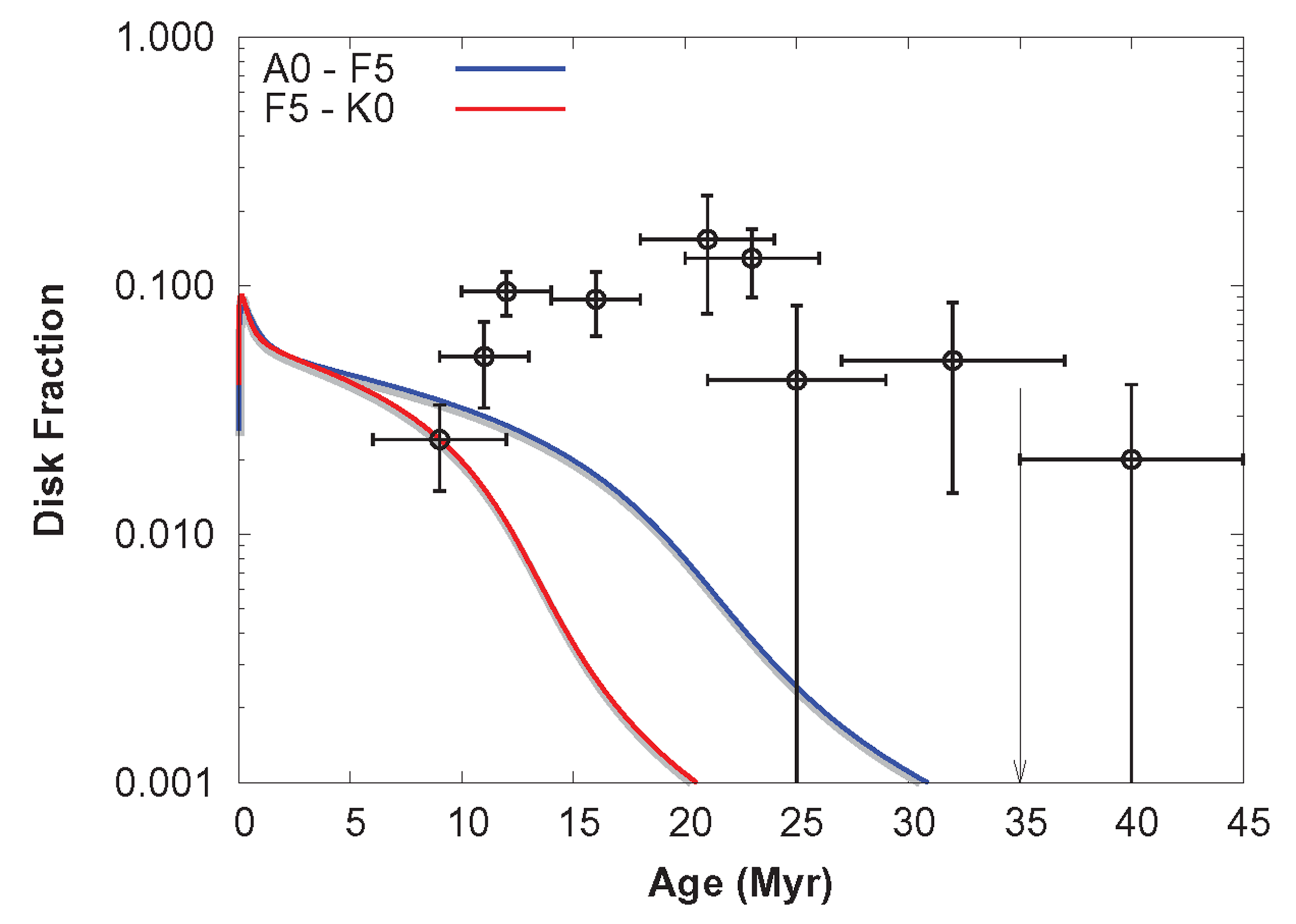}
\caption{The observed fraction of sources with excesses 300\% above the photosphere, with the contamination from protoplanetary disks removed. The blue and red curves show the expected levels of contribution from passively evolving debris disks, based on population synthesis models in \citet{gas13} for early-type and solar-like stars, respectively. The time axis is set to agree with the initiation of the evolution in the previous models \citep{gas13} and the initial evolution should not be taken literally. To fit the ``second component'' peak, the figure suggests the need for high levels of dust production via stochastic events during the oligarchic growth period of planetary formation.\label{model}}
\end{center}
\end{figure}

To test this possibility, we re-analyzed the models in \citet{gas13}. They modeled the 24 $\micron$ excess decay of a sample of 1097 sources with the numerical collisional model CODE-M via population synthesis. By placing the disks at $\sim$5 AU for solar-mass stars and $\sim$11 AU for early-type systems and assuming a log-normal distribution of initial disk masses, they were able to reproduce the observed fraction of disks with $\ge$10\% excesses (i.e., $\epsilon_{lim} = 0.1$). In this work, we use the population of disks synthesized to match the large sample with low detection levels in \citet{gas13}, but set the observational threshold of the population to 300\% above the photosphere (for $\epsilon_{lim} = 3$). This will naturally yield a lower percentage of sources with excesses. In Figure \ref{model}, we show the results of our modeling for the solar-mass and early-type populations, with the disks placed at 5.5 AU from solar-mass stars and 11 AU from early-types. The component from the protoplanetary disks has been removed from the observational data, based on the identifications in the original references, for direct comparison with the models. We have started the cascades at t = 0 as in \citet{gas13}, but with the expanded timescale it would be more appropriate to start them somewhat later, e.g., t = 5 Myr. However, it is clear that no starting time can fit the full population of large excesses. Our analysis rules out steady-state evolution as the primary source of the peak.

\subsubsection{Debris from the Oligarchic/chaotic Phase of Terrestrial Planet Formation}

We conclude that the protoplanetary and transitional disks as well as passively evolving debris disks are a minority of the slowly decaying disks. We now show that these systems are likely to be debris (i.e., second-generation) disks but ones experiencing stochastic events from large impacts that substantially increase their infrared emission.

Many studies \citep[e.g.,][]{kok98,tho03,ken04, ken06, genda15} have suggested that there should be copious amounts of debris generated by stochastic events occurring during the oligarchic/chaotic planetary formation phase. \citet{ken04} first suggested that this emission was likely to peak in the 10 - 40 Myr period (but see \citet{ken16}). \citet{cur08a} found some evidence in support of this suggestion. The self-stirring models of \citet{ken10} also produce broad maxima in the excess emission roughly in this time regime. 

The recent detailed smooth particle hydrodynamics (SPH) simulations of this process by \citet{genda15} substantially improve on these results. They model the growth of terrestrial planets through giant impacts of large protoplanets. Each event results in the injection of a large amount of material (up to $\sim$ 1 M$_\oplus$) into orbit around the star, in the form of small fragments that quickly decay into dusty debris and large fragments that disappear much more slowly. As a result, between 10 and 100 Myr the debris disk around a star has many sharp increases in mass, some of which decay quickly (direct collision products) and some of which decay with characteristic time scales of a few million years (larger collision products that are eroded by collisional cascades). \citet{genda15} compute the 24 $\mu$m emission of the resulting debris system; it reflects this behavior and often exceeds our threshold of three times the photospheric level. The level of emission tends to be largest between 10 and 30 Myr (their Figure 6), in excellent agreement with the results of our study. This possibility is also indicated by Figure 7, if the debris disk evolution is initiated at 10 - 15 Myr rather than at zero. 

\section{Conclusions}

We have analyzed the rate at which 24 $\mu$m excess emission indicative of circumstellar disks fades during the first 40 Myr of evolution, based on a study of 15 stellar clusters and associations. Our study focuses on bright disks, with 24 $\mu$m emission three times or more of the photospheric outputs of the stars. Because of uncertainties in age scales for young clusters/associations, we have carried out the analysis for the traditional scale \citep[e.g.,][]{mam09} and a revised one that proceeds nearly two times more slowly \citep[e.g.,][]{bell13,bell15}. We find that:

\begin{itemize}

\item{On the traditional age scale for young stars, the exponential time scales for disk fading is $3.2 \pm 0.2$ Myr for 24 $\mu$m and $\sim$ 2.5 Myr for wavelengths shorter than 10 $\mu$m \citep{rib14}. On the revised age scale, the time scales 3.5 Myr for 3.6 - 4.5 $\mu$m, 5.3 Myr for 8 - 12 $\mu$m, and $6.6 \pm0.4$ Myr for 24 $\mu$m. }

\item{As indicated by many previous studies, the time scale for fading of protoplanetary disks at 24 $\mu$m is longer than that at shorter wavelengths. }

\item{Based on the behavior of transitional disks, the fading at 24 $\mu$m is not accurately exponential, but is slower than the best fit exponential initially but faster by an age of 10 Myr.}

\item{Taking account of this behavior rather than fitting an exponential indicates a delay of $\sim$ 1 Myr (revised age scale) or less than 1 Myr (traditional scale) in the dispersion of the 24 $\mu$m disk component after the dispersion of the component dominant near 6 $\mu$m, in satisfactory agreement with the theoretical predictions of \citet{gorti15}. This delay also agrees well with the lifetime of roughy 1 Myr for the transitional disk phase. This agreement between observation and theory indicates that disks disperse by photoevaporation and from both the inside and the outside, with the zone dominating the 24 $\mu$m emission being relatively slow to dissipate.}

\item{After this initial fading, there is a continuing bright disk population to $\sim$ 35 Myr.}

\item{This population includes some protoplanetary disks in the 10 - 16 Myr range, but is dominated by bright debris disks.}

\item{After identifying and excluding the contribution from protoplanetary disks, the incidence of debris disks can be described by a log-normal function with the peak on the revised age scale at 18.1 $\pm$ 1.1 Myr where it is 12.3\% $\pm$ 2.9\% of the original population. After the peak, the incidence of debris disks decays with an indicated lifetime of $11.8^{+2.6}_{-3.3}$ Myr. The corresponding values on the traditional scale are 11.8 $\pm$ 1.2 Myr, 13.1\% $\pm$ 3.4\%, and $16.0^{+6.1}_{-4.8}$ Myr.}

\item{These bright debris systems are probably associated with stochastic events caused by large impacts 
during the oligarchic/chaotic phase of terrestrial planet building. The systems we detect appear to correspond well with the theoretical predictions for this process by \citet{genda15}. }

\end{itemize}

\section*{Acknowledgements}
We thank Andrew Youdin for a helpful discussion. This work is based on observations made with the cryogenic mission of the Spitzer Space Telescope, which was operated by the Jet Propulsion Laboratory, California Institute of Technology under a contract with NASA. This publication makes use of data products from the Two Micron All Sky Survey, which is a joint project of the University of Massachusetts and the Infrared Processing and Analysis Center/California Institute of Technology, funded by the National Aeronautics and Space Administration and the National Science Foundation. It also makes use of data products from the Wide-field Infrared Survey Explorer, which is a joint project of the University of California, Los Angeles, and the Jet Propulsion Laboratory/California Institute of Technology, funded by the National Aeronautics and Space Administration. This project was partially supported by contracts 1255094 and 1256424 from Caltech/JPL to the University of Arizona, and by NASA Grant NNX10AD38G.

\facility{Spitzer (MIPS, IRAC), WISE}

\clearpage

\begin{deluxetable}{lllccclcl}
\tabletypesize{\footnotesize}
\tablecaption{Log of MIPS 24 $\micron$ Observations of Young Clusters\label{list}}
\tablehead{
& & \colhead{Start Time} & & \colhead{Sky Annulus\tablenotemark{b}} & & & \colhead{$F_{24,lim}$\tablenotemark{d}} &   \\
\colhead{Target} & \colhead{AOR} & \colhead{(UTC)} & \colhead{         $R$\tablenotemark{a}} & \colhead{$R_i$} & \colhead{$R_o$} & \colhead{$f$\tablenotemark{c}} & \colhead{(mJy)} & \colhead{$N_{p}/N$ (rej\%)\tablenotemark{e}}
}
\startdata
NGC 1333 & 4316672 & Feb 03, 2004, 03:00 & 4 & 4 & 8 & 2.170 & 0.6 & 101/163 (38\%) \\
NGC 1960 & multiple & Oct 23, 2008, 15:28 & 4 & 5 & 10 & 2.124 & 0.3 & 124/128 (3\%) \\
NGC 2232 & 3961856 & Apr 04, 2005, 12:40 & 4 & 15 & 30 & 1.878 & 0.4 & 203/203 (0\%)\tablenotemark{f} \\
NGC 2244 & 4316928 & Mar 15, 2004, 19:06 & 4 & 5 & 10 & 2.124 & 1.1 & 132/369 (64\%) \\
NGC 2362 & 4317440 & Mar 16, 2004, 18:10 & 4 & 15 & 30 & 1.878 & 0.2 & 272/330 (18\%) \\
NGC 2547 & 4318976 & Jan 28, 2004, 19:11 & 4 & 4 & 8 & 2.170 & 0.1\tablenotemark{g} & 219/282 (22\%) \\
& multiple & May 09, 2006, 15:03 & & & & & & \\
& multiple & Dec 01, 2006, 22:02 & & & & & & \\
IC 348 & 4315904 & Feb 21, 2004, 00:13 & 4 & 4 & 8 & 2.170 & 1.5 & 118/339 (65\%) \\
& 22961920 & Sep 23, 2007, 07:49 & & & & & & \\
& 22962176 & Sep 24, 2007, 08:57 & & & & & & \\
& 22962432 & Sep 25, 2007, 08:32 & & & & & & \\
& 22962688 & Sep 26, 2007, 08:58 & & & & & & \\
& 22962944 & Sep 27, 2007, 06:00 & & & & & & \\
& 22963200 & Mar 12, 2008, 14:23 & & & & & & \\
& 22967552 & Mar 18, 2008, 17:28 & & & & & & \\
IC 2395 & 4315392 & Apr 11, 2004, 22:39 & 4 & 4 & 8 & 2.124 & 0.3 & 174/297 (41\%) \\
IC 4665 & multiple & May 19, 2008, 05:33 & 4 & 4 & 8 & 2.124 & 0.1 & 60/60 (0\%)\tablenotemark{f} \\
Ori OB1a & 10987008 & Mar 04, 2005, 03:34 & 4 & 12 & 20 & 1.882 & 0.2 & 112/115 (3\%) \\
Ori OB1b & 10986752 & Mar 03, 2005, 22:25 & 4 & 12 & 20 & 1.882 & 0.4 & 85/106 (20\%) \\
UCL+LCC & multiple & multiple & 4 & 4 & 8 & 2.124 & \nodata & 136/136 (0\%)\tablenotemark{f} \\
Upper Sco & multiple & multiple & 4 & 4 & 8 & 2.124 & \nodata & 237/237 (0\%)\tablenotemark{f} \\
\enddata
\tablenotetext{a}{Photometric aperture radius in pixels.}
\tablenotetext{b}{Inner and outer radii of sky annuli in pixels.}
\tablenotetext{c}{Aperture correction factor.}
\tablenotetext{d}{MIPS 24 $\micron$ detection limit at cluster distances.}
\tablenotetext{e}{Numbers of filter-passed stars ($N_p$) and all member stars ($N$). Filter rejection rates are given in percentage in parentheses. Stars without IRAC 3.6 $\micron$ magnitudes or out of MIPS 24 $\micron$ fields do not count.}
\tablenotetext{f}{No filter applied.}
\tablenotetext{g}{Detection limit in the mosaic image.}
\end{deluxetable}

\clearpage

\begin{deluxetable}{ccccccccc}
\tablecaption{Summary of 24 $\mu$m photometry}
\tablehead{
\colhead{Cluster/}& \colhead{Running} & \colhead{2MASS ID}  &\colhead{RA} & \colhead{Dec} & \colhead{$F_{24}$ }  & \colhead{ $F_{24}$ err } & \colhead{ Treatment\tablenotemark{a}} & \colhead{ $\epsilon$\tablenotemark{b} }\\
\colhead{Assoc.}  &  \colhead{ID}   &          &   \colhead{J2000.0 }   &  \colhead{J2000.0 }   &   \colhead{mJy }    &  \colhead{ mJy}    &      &     
}
\startdata
NGC 1333	&	1	&	2MASS j03281101+3117292	&	52.045899	&	31.291464	&	0.827	&	0.096	&	D	&	0.738	\\															
NGC 1333	&	2	&	2MASS j03281518+3117238	&	52.063279	&	31.289967	&	0.534	&	0.076	&	ND	&	31.654	\\															
NGC 1333	&	3	&	2MASS j03282839+3116273	&	52.118321	&	31.274265	&	-0.138	&	0.017	&	ND	&	-3.807	\\															
NGC 1333	&	4	&		&	52.123917	&	31.255722	&	-0.121	&	0.028	&	ND	&	-39.612	\\															
NGC 1333	&	5	&	2MASS j03283107+3117040	&	52.129498	&	31.284468	&	0.345	&	0.089	&	ND	&	2.826	\\															
NGC 1333	&	6	&	2MASS j03283258+3111040	&	52.135756	&	31.184471	&	56.216	&	0.518	&	MR	&		\\															
NGC 1333	&	7	&	2MASS j03283450+3100510	&	52.143754	&	31.01417	&	1291.387	&	29.735	&	MR	&		\\		
\enddata
\tablenotetext{a}{D = individually detected at 22 - 24 $\mu$m, ND = not individually detected, R = rejected by filter, MR = manually rejected (see text), MA manually added (see text), N = filter not used and all members are detected at 22 - 24 $\mu$m}
\tablenotetext{b}{Nominal excess-to-photospheric flux ratio}
{(This table is available in its entirety in a machine-readable form in the online journal. A portion is shown here for guidance regarding its form and content.)}
\end{deluxetable}

\clearpage



\floattable
\begin{deluxetable}{ccccccccc}
\tabletypesize{\scriptsize}
\rotate
\tablecaption{Fraction of members with excesses above $\epsilon_{lim} = 3$ \label{fracs}}
\tablehead{
\colhead{Cluster/Assoc.}& \colhead{Trad. Age (Myr)*} & \colhead{Rev. Age (Myr)} & \colhead{ fraction} & \colhead{uncertainty} & \colhead{fraction** $0.5 \le M_\odot < 1.2$}  &  \colhead{uncertainty}  &  \colhead{fraction** $\ge$ 1.2 M$_\odot$}  & \colhead{ uncertainty}
}
\startdata
NGC 1333 & 1  & $<$ 6 & 0.765 &  0.100  &  -- & -- & -- &  -- \\
NGC 2244 & $\sim$ 2 & $\sim$ 2 & 0.705  & 0.084 &  -- & -- & -- &  -- \\
Taurus      & 1.5 & 3.5 &  0.749 &  0.057 &  -- & -- & -- &  --\\
$\rho$ Oph  & 1-2 & 4.5  &  0.462  &  0.060 &  -- & -- & -- &  -- \\
IC 348 & 2.5 & 6 & 0.585 & 0.079 & -- & -- & -- &  -- \\
Cha I   & 2.6 & 6  &  0.617  &  0.073  & -- & -- & -- &  -- \\
IC 2395 & 6 & 9 &  0.205 & 0.046 &  0.152 & 0.080  & 0.148  &  0.049  \\
Ori OB1b & 4-6 & 9 & 0.121 & 0.049 & 0.182  &  0.091  &  0.176  &  0.102  \\
Upper Sco & 5 & 11  &  0.143  &  0.025 &  0.119  &  0.053  & 0.096  &  0.027  \\
NGC 2362 & 5 & 12  &  0.137 & 0.044 &  --  &  --  & 0.137  &  0.044  \\
Ori OB1a & 7-10 & 15  & 0.092 & 0.031  & 0.042  &  0.042  & 0.150  &  0.087 \\
UCL+LCC & 16 & 16 &  0.103 & 0.028 & *** & -- &  0.103  &  0.028 \\
$\beta$ Pic MG  & 21 &  21  &  0.103  &  0.042 & 0.080  &  0.057 & 0.192  &  0.086 \\
NGC 1960  & 23 &  23  & 0.129  &  0.040  & -- & -- &  0.129  &  0.040 \\
IC 4665  & 25 & 25  &  0.033  &  0.024  &  0.0  &  0.045 & 0.042  &  0.042 \\
NGC 2232  & 32 &  32  &  0.035  &  0.015  & 0.027 & 0.014  & 0.050  &  0.035 \\
NGC 2547  & 35 & 35  &  0.009  &  0.006  &  0.010 & 0.007& 0.000  &  0.038 \\
Moving Groups  & 40 &  40  &  0.012  &  0.012  &  0.0 & 0.034 &0.020  &  0.020 \\
Pleiades  &  126  & 126 & 0.014  &  0.014  &  -- & -- & 0.034  &  0.034 \\ 
\enddata
\tablenotetext{*}{References and discussion of age assignments can be found in \citet{bal16}.}
\tablenotetext{**}{   Shown for clusters of age $\ge$ 9 Myr used to calculate debris disk  behavior. See Section 4.7.}
\tablenotetext{***}{ ~~~~~~~     Membership list does not extend to 0.5 M$_\odot$.}
\end{deluxetable}

\clearpage



\appendix

\section{Cluster and Association Membership and Ages}\label{cluster_list}

{\bf NGC 1333:} is one of the youngest clusters in our sample with a number of Class 0 protostars, some associated with stellar outflows \citep{wal08}. As part of the Per OB2 association, the cluster is embedded in the western edge of the Persus Cloud \citep{bal08}, an active site of star formation at 240 pc \citep{hir08}. NGC 1333 has been a target of star formation research in a wide range of wavelengths, from X-ray \citep[e.g.,][]{get02} to radio \citep[e.g.,][]{kne00}. The age of the cluster is estimated to be $\sim$2 Myr \citep{flah08, win09} (traditional scale) or $\le$ 6 Myr (revised scale, see \citet{bal16}). We used the membership list published by \citet{luh16}. The 24 $\mu$m photospheric fluxes are based on the IR color of young photospheres in  \citet{luh10} and derived from the J magnitudes of the members at which wavelength an excess is most unlikely. Prior to the availability of this list, we had compiled our own list from the {\it Spitzer} observations of this cluster presented by \citet{gut08} plus the {\it Chandra} observation described in \citet{win10}. This list included some stars identified as members purely on the basis of infrared properties in the IRAC bands; such sources are excluded from the list by \citet{luh16}. The qualified disk fraction in the older calculation was 77.0\% $\pm$ 10.0\%, while the new calculation obtained 67.2\% $\pm$ 10.7\% considering only the member stars with spectral types of M6 or earlier (to make the list comparable among the full set of clusters). Because the number of qualified excesses is high, the difference in these two calculations is not large and would have no effect on our conclusions. A similar situation will obtain for all the young clusters, where membership identification is complex because of extinction. That is, making a valid identification of sources with excesses within the adopted membership list is important, but within this constraint our study is relatively robust against different approaches to identifying the members of the young clusters. 

{\bf NGC 1960:} At a distance of 1.32 kpc, the age determined for NGC 1960 ranges from $\sim$16 Myr \citep{san00}, $\sim$20 Myr \citep{may08,bell13}, to $\sim$25 Myr \citep{sha06}. Since there is no significant difference between the two age scales for clusters older than 15 Myr, we adopt the age of 23 Myr from \citet{sod14} based on the lithium depletion boundary, consistent with other independent estimates \citep{jef13} (we use this single age scale for all the clusters in this age range). The cluster membership list used in this work, including 132 stars, is the combination of the two lists in \citet{smi12}. The membership identification is based on optical color, proper motion, lithium abundance, and radial velocity. However, two members \citep[2\_195 and 2\_277 in][]{smi12} are off the MIPS image and excluded from this analysis.

{\bf NGC 2232:} NGC 2232 is a young cluster in the Gould Belt. Independent papers suggest an age of 25 Myr \citep{lyr06} or 32 Myr \citep{sil14}, and a distance of $\sim$350 pc. We use the list of 209 members of the cluster, identified with X-ray activity, spectroscopy, proper motion, and optical and near-infrared colors \citep{cur08b}.

{\bf NGC 2244:} NGC 2244 is the well-studied core open cluster in the Mon OB2 association, located in the Rosette Nebula 1.4 kpc away from the sun \citep{hen00}, and with a traditional age of $\sim$2 Myr \citep{hen00,par02}. This value agrees with the revised estimate \citep{bell13}, and we adopt it. We constructed a membership list based on X-ray detection, IR excess, proper motion, optical color selection, and a cut-off in the clustrocentric distance. The details are given in Appendix~\ref{mem_list}.

{\bf NGC 2362:} At a distance of 1.48 kpc \citep{moi01}, \objectname{NGC 2362} is centered on the O9 Ib star \objectname{$\tau$ CMa} and is $\sim$3 pc in radius \citep{dah07}. The age is $\sim$5 Myr on the traditional scale \citep{bal96,moi01,dah05}, and on  the revised scale is 12 Myr \citep{bell13}. We combined the two member lists in \citet{dah07}, based in the first case on X-ray detection, lithium abundance, and H$\alpha$ emission and in the second on optical photometry, for a total of 330 stars.

{\bf NGC 2547:} This is a well investigated cluster. The X-ray observations, main sequence turn-off, lithium abundance, and optical color together suggest an age of $\sim$38 Myr, and a corresponding distance of $\sim$360 pc \citep[][and the references therein]{gor07}. \citet{sod14} find an age of 35 Myr from the lithium depletion boundary. We adopt Table 1 in \citet{gor07} for highly probable photometric members, but exclude all stars with no proper motion measurements or with the measured proper motions outside 2-$\sigma$ from that for the cluster. Finally, there are 311 stars left in our final membership list. This number agrees very well with the member selections by \citet{irw08}, who found $\sim$800 photometric member candidates of \objectname{NGC 2547} in a comparable sized field of view, and expected $\sim$330 of them to be real members based on simulations. Therefore, we estimate our membership list to be practically free of contamination. The photometry in this work is based on the mosaic image of all observations \citep{gor07,for08}.

{\bf IC 348:} The open cluster \objectname{IC 348}, at a distance of $\sim$ 300 pc, is rich in late type stars \citep[see][for a general review]{herb08}. Traditionally thought to have an age of $\sim$2.5 Myr, \citet{bell13} revise this value to be as old as $\sim$ 6 Myr. As might be expected from its severe and highly variable extinction in the visible regime, \objectname{IC 348} has a complicated nebulous background at 24 $\micron$. We took members from \citet{lad06} and \citet{mue07}, excluding Table 2 in \citet{mue07} that may include interlopers. The list includes 339 stars, and is expected to be largely unbiased since a number of criteria are employed, including proper motion, lithium abundance, surface gravity, photometric extinction, and emission lines. We have compared this list with the one recently published by \citet{luh16} and find virtually complete agreement except for a much larger number of late M and brown dwarfs in the newer list; these are excluded from our analysis in general to make the lists comparable among the full set of clusters.

{\bf IC 2395:} The galactic cluster IC 2395 is $\sim$800 pc from the sun and its age was previously estimated to be $\sim$6 Myr \citep{cla03}; on the revised age scale, it is about 9 Myr \citep{bal16}. The membership list of 280 stars, and a study of the excesses in this cluster, are discussed in \citet{bal16}. Members have been identified through proper motion and radial velocities, in addition to photometry.

{\bf IC 4665:} A cluster at $\sim$370 pc, IC 4665 has an age of 25 Myr indicated by its lithium depletion boundary \citep{man08,sod14}. \citet{smi11} have a thorough membership list based on radial velocity, proper motion, spectroscopy, and optical color.

{\bf Chamaeleon I:} For Chamaeleon I, we adopt the membership list from \citet{luh08_review} based primarily on spectroscopy and the {\it Spitzer}/IRAC and MIPS photometry from Table 7 in \citet{luh08}. For our purpose of disk census at 24 $\micron$, only stars with both 3.6 and 24 $\micron$ photometry were retained. We adopt $6 \pm 1$ Myr as the revised age, as Chamaeleon I members are nearly indistinguishable in color-magnitude diagrams with members of IC 348 and $\sigma$ Ori \citep{luh07, luh10}, while the latter two are 6 Myr old based on the pre-main sequence isochrone model \citep{bell13}. On the traditional scale, an age of $\sim$ 2.6 Myr has been estimated \citep{luh08}.

{\bf Orion OB1:} Orion OB1 is a nearby OB association with a clear age sequence. In this work, we consider two of its subgroups separately: Ori OB1a and OB1b. The former, around the 25 Ori cluster, appears older and closer to us at 7-10 Myr and $\sim$330 pc; while the latter is younger and farther with an age of 4-6 Myr at $\sim$440 pc on the traditonal scale \citep{bri05,bri07}. The revised ages are $\sim$ 15 Myr for OB1a and $\sim$ 9 Myr for OB1b \citep[see discussion in][]{bal16}. \citet{bri05,bri07} provided good membership identifications for both subassociations, based on optical and infrared color, spectroscopy including lithium depletion and H$\alpha$ emission, and radial velocity. We used these membership lists with 115 confirmed members in OB1a and 106 in OB1b. \citet{hern07b} gave the optical photometry for 106 additional member candidates in both subassociations. However, as they estimated that $\sim$50\% of the candidates could be interlopers in OB1b, we discard these candidate members altogether to ensure contamination-free membership lists.

{\bf Taurus star-forming region:} The Taurus star-forming region is one of the closest to the sun of its kind \citep{ken08_taurus}. In this work, we take the membership list in \citet{luh10} with existing IRAC and MIPS photometry in their Table 4 and 6, respectively. For the stars with photometry at multiple epochs, we adopt the average. This accounts for all relevant {\it Spitzer} observations during the cryogenic mission. Our adopted age of 3.5 Myr \citep{rees15} on the revised scale  is consistent within the errors with the estimate of 2.5 Myr from surface gravity dating of 10 solar-mass pre-main sequence stars \citep{tak14}. The traditional age estimate is 1.5 Myr \citep{bar03}.

{\bf Tucana-Horologium, Columba, Carina, and Argus Associations:} These are largely independent stellar associations defined based on their common spatial motion, age, and distance \citep[e.g.,][]{zuc04,tor08}. Here we combine them as one group because of their similar ages of about 40 Myr \citep{zuc00,tor08}. The membership lists used in this work are taken from Table 3 in \citet{mal13}, which are based on Bayesian analysis of the kinematics and the $I_C$- and $J$-band magnitude/color of the stars, and should be unbiased towards disk presence.

{\bf Upper Scorpius:} Upper Sco is the youngest group in the \objectname{Sco-Cen association} at an average distance of $\sim$140 pc. \objectname{Upper Sco}, with a traditional age of 5 Myr old and a revised value of $\sim$11 Myr \citep{pec12}. MIPS can reach the stellar photospheres of all the member stars at 24 $\micron$ down to late types \citep[$\sim$M5,][]{car09}. The membership list we use has 237 stars, consisting of 220 member stars in \citet{car09}, supplemented by 17 stars in \citet{che11}. The members are selected based on proper motion, optical colors and X-ray activity, and should be unbiased with respect to disk presence \citep{car09}.

{\bf Upper Centaurus-Lupus and Lower Centaurus-Crux:} UCL and LCC are the other two major subassociations in Sco-Cen. Their age estimates are $\sim$15 and 17 Myr, respectively \citep{mam02}, both significantly older than Upper Sco. Given their identities in the same OB association and the similar ages, here we treat them like one single group. We use the membership list in \citet{che11} that only focuses on F- and G-type stars. The membership identifications are based on proper motion, radial velocity, and spectroscopy and lithium abundance.

{\bf $\beta$ Pictoris Moving Group:} The BPMG is a group of young stars close to the sun with consistent spatial motion. The age was estimated to be $\sim$12 Myr \citep{zuc04}, but \citet{sod14} argue from the lithium depletion boundary that 21 Myr is more likely, consistent with the result of \citet{bin14}, which we adopt. We drew our membership list from the literature \citep{zuc04,tor06,reb08,tor08,das09,lep09,via09,ric10,sch10,kis11}. However, we do not use the members identified by \citet{moo06}, because they used infrared excess as the selection criterion so the results could be biased towards disk-bearing stars. The final list includes 63 members, where the secondaries in resolved binary or multiple systems are not counted. These BPMG stars are not all covered by MIPS data. For internal consistency, we use {\it WISE} W4 \citep[effective wavelength 22 $\micron$,][]{jar11} as the substitute to take advantage of the proximity of the group. Given its similar wavelength and spectral response, the results from {\it WISE} W4 should be directly comparable with the MIPS 24 $\micron$ outcomes. We adopt the W4 photometry in the AllWISE Source Catalog for most sources, but also carry out aperture photometry for stars with saturated pixels or $\chi^2 > 3$ in PSF-fitting (non-point sources, presumably with resolved disks). The only exception is HD 322990, which is not bright enough at W4 to be detected by WISE. Fortunately, this star was detected in a Spitzer/MIPS 24 $\mu$m Galactic plane survey image (AOR 20500480). We use the MIPS 24 $\mu$m photometry as a substitute for the missing WISE W4 detection.

{\bf $\rho$ Ophiuchi star-forming region:} At $\sim$120 pc, $\rho$ Oph is one of the nearest star-forming regions and is well studied \citep{wil08}. Many newborn stars near the young core region are deeply embedded in dark nebulae with severe extinction \citep{eva03}. In this work, we adopt an unbiased, extinction-limited membership list of 135 stars in the outer fringe population from Table 3 in \citet{eri11}, identifed through optical spectroscopy. The corresponding {\it Spitzer} photometry is pulled from the c2d catalog \citep{eva03,eva09} through VizieR. We retain all 130 stars that have MIPS 24 $\micron$ photometry. Most (123) of these stars also have robust photometry at IRAC 3.6 $\micron$, which we use to derive the photospheric output at 24 $\micron$. For those that are not observed at 3.6 $\micron$ by IRAC or have poor photometry, we use 2MASS $K_s$ photometry as the substitute by considering the differential extinction between $K_s$ and 24 $\micron$ \citep{fla07}. This population is slightly older than the embedded star formation core with an estimated average age of 3.1 Myr on the traditional age scale \citep{eri11}. In this work, we adopt an average age of $4.5 \pm 1.5$ Myr on the revised scale.

\section{Membership Identification for NGC 2244}\label{mem_list}


For NGC 2244, we assembled a list of member candidates from the {\it Chandra} X-ray survey \citep{wan08}, complemented by {\it Spitzer}/IRAC observations \citep{bal07}. First, we merged all sources detected in either X-rays or with IRAC into one catalog, which is then matched to the 2MASS catalog \citep[2MASS,][]{skr06} for $JHK_S$ counterparts and accurate positions. We allowed a radius of 2" in all coordinate matches to accommodate astrometric errors. For stars with more than one candidate identification, we selected the closest; if there was no 2MASS counterpart, the IRAC position \citep{bal07} was adopted. All stars detected in both the infrared and X-ray are considered as viable member candidates. We extrapolated the IRAC 3.6 $\micron$ flux densities to 8.0 and 24 $\micron$ via the Rayleigh-Jeans approximation. A source was also considered a cluster member candidate if its flux was at least twice the extrapolated photospheric value in at least one of the two wavebands.

The member candidates were then screened for appropriate photometric properties and proper motions. \citet{ogu81} obtained optical photometry in \objectname{NGC 2244} down to $V \approx 14$ and found $\sim$ 170 members within a total of $\sim$ 400 stars. We rejected all candidate members that were questionable from the optical photometry. \citet{mar82} used proper motion measurements on photographic plates taken from 1914 to 1974 to determine membership probabilities. \citet{sab01} applied an improved proper motion distribution model to the best measurements (quality flag 9 or 10) in the same data; we use these results. The UCAC2 catalog \citep{zac04} also covered the field of \objectname{NGC 2244} with proper motion measurements, which were utilized by \citet{dia06} for membership probabilities. Finally, membership probabilities were derived by \citet{che07} with independent observations of the field down to $B \approx$ 16 from 1963 to 1999. The membership probabilities obtained in different proper motion studies agree with each other for many of the stars, but there are also discrepancies. We required our final ``cluster members'' to have membership probability $> 50\%$ in all studies. Objects that failed these tests were still retained if they fall within 9\arcmin\ of the cluster center at R.A. = 6$^\textrm{h}$31$^\textrm{m}$58$^\textrm{s}$ and Dec. = +4$\arcdeg$54$\arcmin$51$\arcsec$, where there is a concentration of Class II objects found by \citet{bal07} with IRAC observations. This position is in good agreement with the center found in X-ray \citep{wan08} and 2MASS $JHK_S$ bands \citep{li05}. The distance cutoff at 9\arcmin\ is consistent with the radial profile of the cluster \citep{li05}. Our final membership list in Table~\ref{member_ngc2244} includes 718 stars, of which 369 have IRAC detections \citep{bal07}. We plot the member and non-member stars in Figure~\ref{ngc2244field}, and compare their surface number densities in Figure~\ref{surden}. As expected, the surface density of member stars decays with the clustrocentric distance, whereas that of non-member stars remains nearly constant at a low level.

\clearpage

\setcounter{figure}{0} \renewcommand{\thefigure}{B.\arabic{figure}}

\begin{figure}
\center
\includegraphics[angle=+0,width=12cm]{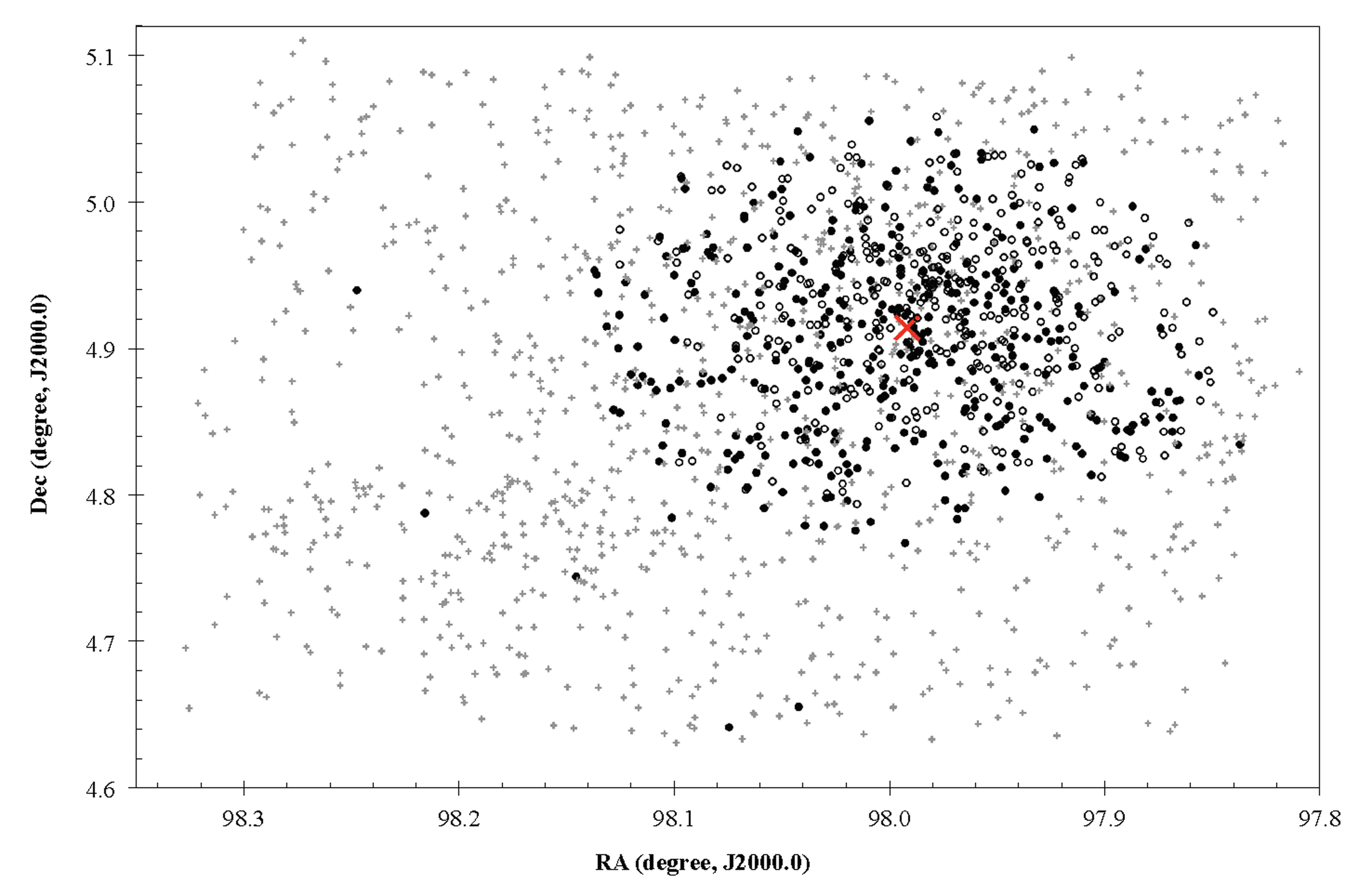}
\caption{Field of NGC 2244. Cluster members with and without IRAC 3.6 $\micron$ detection are represented by filled and open circles, respectively. The large cross labels the cluster center mentioned in the text. Grey pluses are non-member stars.\label{ngc2244field}}
\end{figure}

\clearpage

\begin{figure}
\center
\includegraphics[angle=+0,width=12cm]{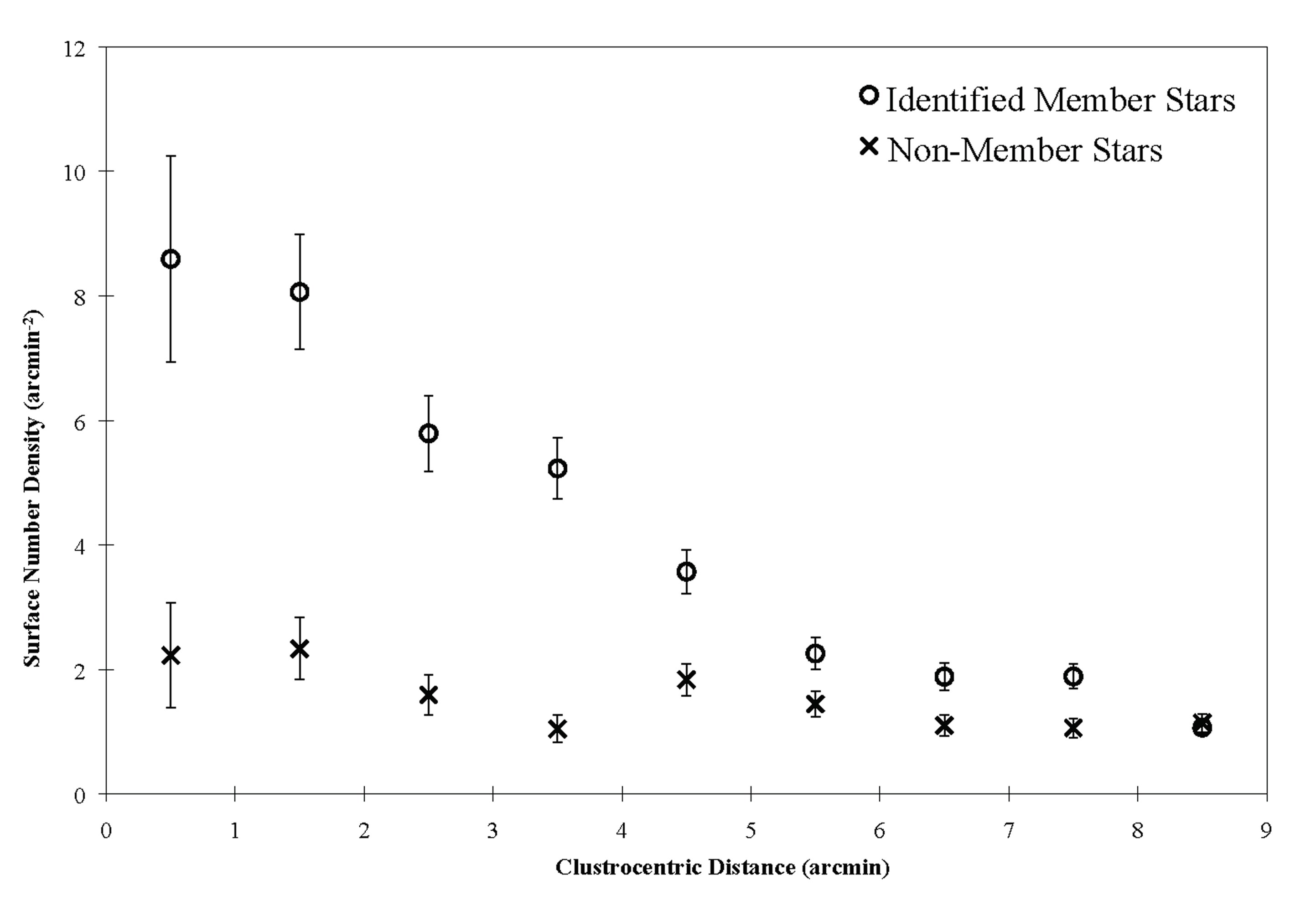}
\caption{Surface density of identified members and non-members of NGC 2244. The error bars represent 1 $\sigma$ values calculated with $\sqrt{N}$.
\label{surden}}
\end{figure}

\clearpage

\clearpage

\begin{deluxetable}{lccccc}
\tabletypesize{\small}
\tablecaption{Final Membership List of NGC 2244\label{member_ngc2244}}
\tablewidth{0pt}
\tablehead{
ID\tablenotemark{a} & Catalog ID\tablenotemark{b} & R.A. (deg) (J2000.0) & Decl. (deg) (J2000.0) & $A_{K_s}$\tablenotemark{c} & Cross-link\tablenotemark{d}
}	
\startdata
1 & W8 / B16 & 97.836969 & 4.834407 & 0.16 & M84=C9=OI84 \\
2 & W26 / B42 & 97.856241 & 4.881617 & 0.23 & \\
3 & W27 / B43 & 97.857540 & 4.970587 & 0.60 & PS10 \\
\vdots & \vdots & \vdots & \vdots & \vdots & \\
367 & B821 & 98.145636 & 4.744226 & 0.38 & M279=C67=OI279 \\
368 & W899 / B959 & 98.215799 & 4.787823 & 0.26 & M334=C205=OI334 \\
369 & B992 & 98.247407 & 4.939592 & 0.15 & M354=OI353 \\
\hline
\nodata & W16 & 97.849656 & 4.924812 & \nodata & \\
\nodata & W18 & 97.850911 & 4.877090 & \nodata & \\
\nodata & W19 & 97.851761 & 4.884842 & \nodata & \\
\vdots & \vdots & \vdots & \vdots & \vdots & \\
\nodata & W793 & 98.107374 & 4.973113 & \nodata & \\
\nodata & W820 & 98.125221 & 4.957283 & \nodata & \\
\nodata & W821 & 98.125333 & 4.981250 & \nodata & \\
\enddata
\tablenotetext{a}{For member stars with IRAC detections. Members with X-ray detections only are listed at the bottom without this running ID.}
\tablenotetext{b}{Star sequence in B = \citet{bal07}, W = \citet{wan08}.}
\tablenotetext{c}{Total extinction in 2MASS $K_s$ band.}
\tablenotetext{d}{Star sequence in M = \citet{mar82}, D = \citet{dia06}, C = \citet{che07}, OI = \citet{ogu81}, PS = \citet{par02}}
{(This table is available in its entirety in a machine-readable form in the online journal. A portion is shown here for guidance regarding its form and content.)}
\end{deluxetable}

\clearpage


\section{Filtering out Bad Photometry on Complicated Sky Backgrounds}\label{filter}

Many young open clusters are superimposed on highly nonuniform sky backgrounds at 24 $\micron$. Visual inspection is usually employed for identification of sources where the background is well-enough behaved that reliable photometry can be obtained. However, this approach favors bright sources that stand out against the background structure, and therefore is biased toward high mass stars and stars with large 24 $\micron$ excesses. To obtain an unbiased sampling of the 24 $\micron$ behavior of members of a cluster requires that the bad photometry be filtered out in a way that does not bias the sample; that is, on the basis of the background behavior alone, with no influence from the properties of the star. This appendix discusses our approach to this challenge.

We have developed a digital filter to reject unreliable photometry. The concept is similar to the filter used by \citet{meg12}, with some additional criteria to make the screening more selective. We take \objectname{IC 348}, a cluster with a complicated background, to test and tune the filter. Several metrics describing the sky pixel statistics were tested; in the end, we settled on three of them. For each of the cluster members, we compute the normalized value for the following three parameters. The first is the average sky brightness, determined by computing the intensity-weighted means of the sky pixel values in sky annuli around a source. We used different sky annuli for different clusters and the choices are listed in Table 1. The second is the skewness of the distribution of the sky pixel values, defined within the sky annuli around the sources. A well behaved background region should show a normal distribution of noise. The third is the gradient of the local sky brightness in a star's surrounding area, measured for convenience on the basis of the signals in four small (4 pixels in radius) circular regions surrounding the star (ten pixels distant in each of the cardinal directions).
 
The three parameters are expected to be mutually independent. Therefore, we make each of them a dimension within a three dimensional Cartesian parameter space.  Since each of the three dimensions has a different unit, we define a scalar threshold in ``distance'' in the parameter space by dividing each by its median for a particular cluster (we do not use the means because they are more easily affected by a relatively small number of extreme values). To evaluate the reliability of the photometry of individual stars, we define a virtual distance in the parameter space as
\begin{equation}\label{dist}
D = \sqrt{ \displaystyle\sum_{i=1}^3 d_i^2},
\end{equation}
where $d_i$ is the $i$-th dimensionless parameter. Stars with relatively small values of $D$ are on clean and well-behaved backgrounds and should have the most reliable photometry. In few cases, one of the parameters has a much wider distribution after normalization, therefore, it dominates the D space. We then adjust the parameter by multiplying all values in that dimension by a suitable constant (see Figure C.1 for an example). 

We calibrated this filter on a sample of stars in \objectname{IC 348} that were first evaluated visually to set plausible image detection limits. We obtained aperture photometry for the cluster and looked for stars measured only at a low ratio of signal to noise ($<$ 8 in the case of \objectname{IC 348}; clusters with cleaner backgrounds allowed lower signal to noise thresholds) but that were measured to be brighter than the image detection limits. A properly tuned digital filter should reject these cases. We defined the cutoff threshold in $D$ at the largest value yielding none of these cases, i.e. examples of stars with poor signal to noise but apparent signals above the image detection limit. Figure~\ref{filtering} shows the distribution of sources below the signal to noise threshold (8) in \objectname{IC 348} in each of the three filter parameters. Only stars with $D$ below the threshold above which false detections began to occur were accepted into our study. We confirmed the operation of the filter by manually inspecting each star that passed its test.

A problem with the filter is that it occasionally rejects very bright stars where the stellar PSFs extend at significant levels to the sky annuli and raise the mean brightness therein. Such stars are rare ($<$ 2\%) of the accepted stars in general; we recover them manually. The recovered stars are selected carefully to minimize the risk of introducing any bias: only very bright stars ($\geq$ 15 mJy) on low and clean background with non-rejected stars nearby are recovered. Since the number of such stars is much fewer than the filtered population in each cluster, the manually recovered stars do not significantly influence our conclusions even if some of them were recovered inappropriately.

Since the parameter space and distance cutoff threshold are defined only on the brightness and distribution of sky background, our method should be neither subject to arbitrary human visual selection, nor biased by any source properties. Hence, it only scales down the sample size for each cluster, but should introduce no selection effects. 


We ran the following simulation to demonstrate the validity of the filtering with a complex field background. We based this simulation on the epoch 1 image of IC 348. Since IC 348 has the most complicated 24 $\micron$ background in all the clusters analyzed in this paper, it provides a worst-case test. The simulation gives us confidence in both the photometry of individual sources and also the ability of the filter to identify the sources where it is acceptable to consider the noise (positive and negative) to be unbiased. 

We made five runs of simulations. For each run, we generated a group of log-normally distributed random numbers as the 24 $\micron$ flux densities of the simulated fake stars. Each fake star was place at random image coordinates. The PSF for the fake stars was the same one that was used to extract our PSF-fitting photometry. The field of IC 348 is not crowded by detected stars at 24 $\micron$. This leads us to avoid adding too many fake stars in a single run, because that would make a larger number of overlapped PSFs, and potentially induce larger photometric error than in the original field. After a few tests, we finally chose 200 fake stars for each run, making the total number of test fake stars 1000. We reduce and analyze the simulated image of each run like a new cluster by applying identical procedures as for the real cluster. This includes taking the simulated fake star coordinates as the initial guess for star positions, allowing a maximum shift of 2\arcsec\ for source matching, making PSF-fitting photometry, applying the filter, and finally visually inspect the image. Finally, 610 stars pass the filtering process.


After going through the entire data reduction procedure, our photometry was compared against the brightness of the fake stars. Since the photometric errors may be correlated with the brightness of the stars, the comparisons were evaluated in units of $\sigma$ (the nominal photometric errors), rather than in units of flux density. The outcomes are shown in Figure~\ref{error}. For the unfiltered case, the wide distribution in units of $\sigma$ indicates that the photometric errors have been statistically underestimated, probably because the algorithm for error estimation did not work well with such complicated nebulous background. Nevertheless, we find a significant asymmetry in the photometric error distribution in the original 1000 fake stars. The mean bias is +2.2-$\sigma$ brighter than the input values, with a skewness of 14.1. However, after applying the filter, the retained population of 610 stars has a mean bias of +0.4-$\sigma$ and a skewness of 0.8, which is essentially symmetric with respect to 0. 

In addition, this result supports our assumption of symmetry in the noise, allowing the extraction of excess positive signals on a statistical basis even if they are below the threshold for individual detections. This conclusion holds because we applied this test at the positions of cluster members identified at shorter wavelengths, but only if they passed the screening for well-behaved background using the digital filter. 

\clearpage

\setcounter{figure}{0} \renewcommand{\thefigure}{C.\arabic{figure}}

\begin{figure}
\center
\includegraphics[angle=+0,width=12cm]{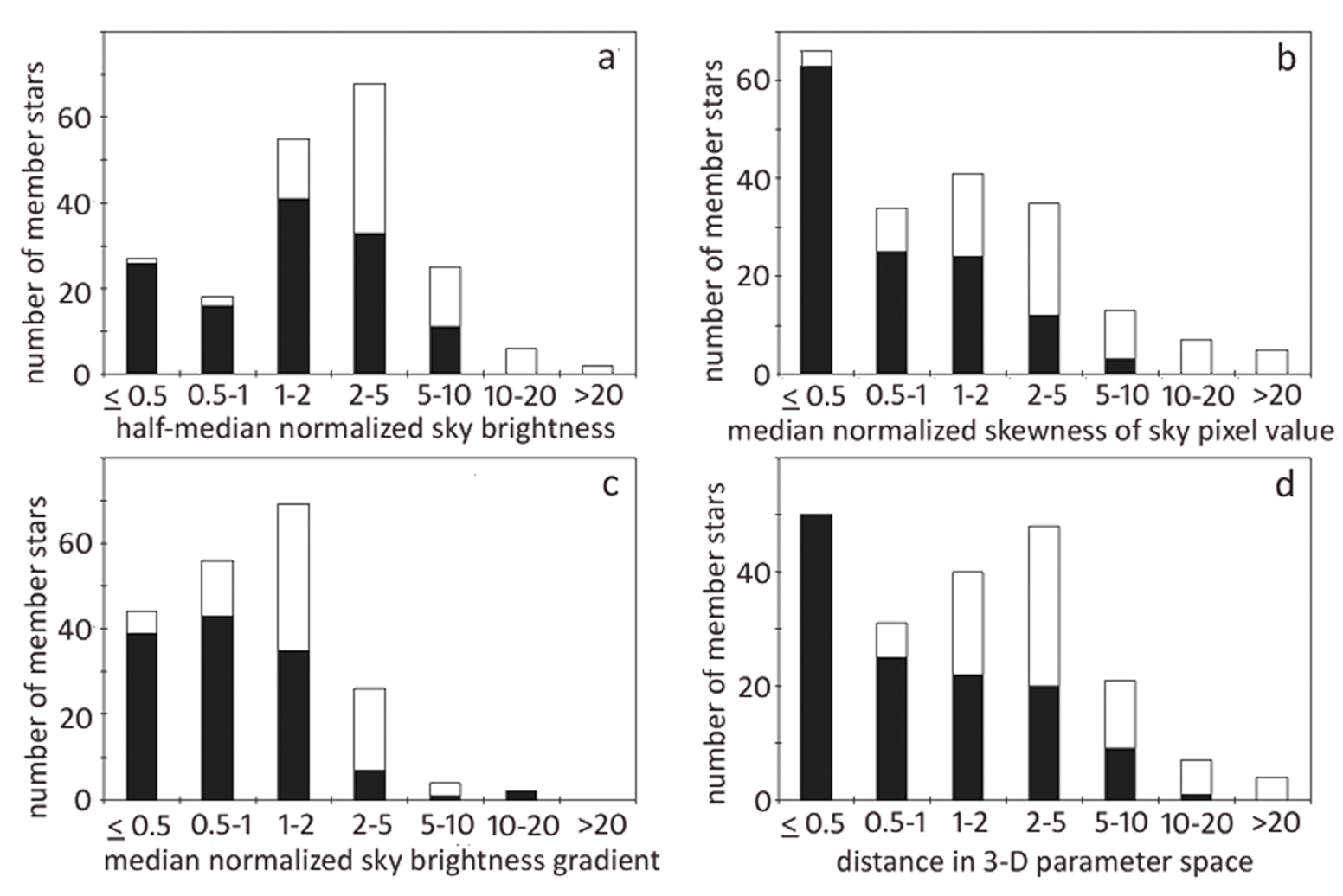}
\caption{Distribution of the filter parameters and the ``distance'' of the stars in IC 348, epoch 1. The filled histogram shows the numbers of stars with measured fluxes less than the detection limit, while the open one is for cases where the measured flux is $>$ the detection limit. With the secondary sample, measured flux greater than the detection limit means unreliable photometry. Panels are for the distribution of a) normalized sky brightness; b) normalized skewness of sky pixel values; c) normalized sky background gradient; and d) distance in the parameter space. The values of the normalized sky brightness are doubled  by introducing a coefficient of 0.5 to the original distribution median to make it a comparable range with the other two parameters. For valid photometry, the cutoff threshold in distance is 1.85 for IC 348 (stars with distance $> 1.85$ are rejected in panel d).
\label{filtering}}
\end{figure}

\clearpage

\begin{figure}
\center
\includegraphics[angle=+0,width=10cm]{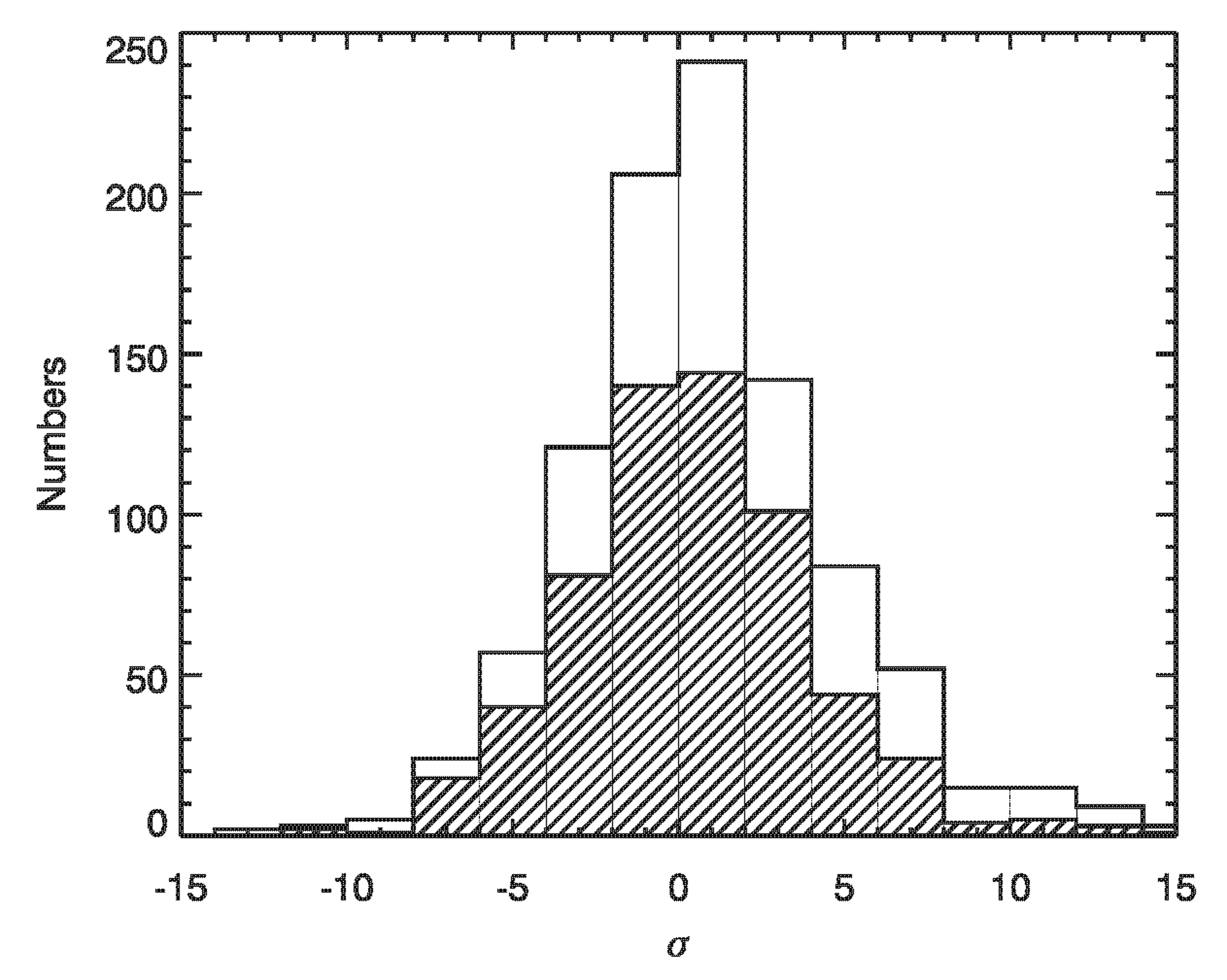}
\caption{Distribution of errors, in units of $\sigma$, from comparing the extracted values with the input ones in the fake star simulation. The distribution for the unfiltered sample (full box heights) is significantly skewed toward the positive side, particularly apparent above 5 $\sigma$. The unfiltered sample also has some cases falling outside the plotted range. However, the filtered population (overplotted shadowed areas) has a near-symmetric distribution. For this test, $\sigma$ was computed on a per-pixel basis; since the data subsample the native pixels into four equivalent ones, the nominal errors are half the values that would apply to photometry and hence the plotted distributions are indicated at twice as many $\sigma$ as would be found for standard deviations applied to photometry. 
\label{error}}
\end{figure}

\end{document}